\documentclass[useAMS,usenatbib]{mn2e}
\usepackage{times} 
\usepackage{listings}
\usepackage{graphics}
\usepackage{subfigure}
\usepackage{graphicx}
\usepackage{graphicx,xspace}
\usepackage{amsmath}
\usepackage{amssymb}
\usepackage{hyperref}
\usepackage{aas_macros}
\usepackage{lscape}
\usepackage{rotating}
\usepackage{placeins}
\usepackage{natbib}
\usepackage{color}

\graphicspath{{./Plots/}}
\pdfpageattr {/Group << /S /Transparency /I true /CS /DeviceRGB>>}

\newcommand{\galform}{{\sc{galform}}\xspace}
\newcommand{\grasil}{{\sc{grasil}}\xspace}

\def\apjl{ApJL}

\def\aap{A\&A}

\def\apjs{ApJS}

\title[Simulated observations of SMGs]{Simulated observations of sub-millimetre galaxies: the impact of single-dish resolution and field variance}
\author[W. I. Cowley et al.]{
\parbox[t]{\textwidth}{
William I. Cowley\thanks{E-mail: w.i.cowley@durham.ac.uk }, 
Cedric G. Lacey,
Carlton M. Baugh,
Shaun Cole}
\vspace*{6pt}\\
Institute for Computational Cosmology, Department of Physics,
University of Durham, South Road, Durham, DH1 3LE, UK.
\vspace*{-0.5cm}}

\begin{document}

\date{\today}

\pagerange{\pageref{firstpage}--\pageref{lastpage}} \pubyear{2014}

\maketitle

\label{firstpage}

\begin{abstract}
Recent observational evidence suggests that the coarse angular resolution ($\sim20''$ FWHM) of single-dish telescopes at sub-mm wavelengths has biased the observed galaxy number counts by blending together the sub-mm emission from multiple sub-mm galaxies (SMGs). We use lightcones computed from an updated implementation of the \galform semi-analytic model to generate $50$ mock sub-mm surveys of $0.5$ deg$^2$ at $850$ $\mu$m, taking into account the effects of the finite single-dish beam in a more accurate way than has been done previously. We find that blending of SMGs does lead to an enhancement of source extracted number counts at bright fluxes ($S_{\mathrm{850}\mu\mathrm{m}}\gtrsim1$ mJy).  Typically, $\sim3{-}6$ galaxies contribute $90\%$ of the flux of an $S_{850\mu\mathrm{m}}=5$ mJy source and these blended galaxies are physically unassociated.  We find that field-to-field variations are comparable to Poisson fluctuations for our $S_{850\mu\mathrm{m}}>5$ mJy SMG population, which has a median redshift $z_{50}=2.0$, but are greater than Poisson for the $S_{850\mu\mathrm{m}}>1$ mJy population ($z_{50}=2.8$). In a detailed comparison to a recent interferometric survey targeted at single-dish detected sources, we reproduce the difference between single-dish and interferometer number counts and find a median redshift ($z_{50}=2.5$) in excellent agreement with the observed value ($z_{50}=2.5\pm 0.2$).  We also present predictions for single-dish survey number counts at $450$ and $1100$ $\mu$m, which show good agreement with observational data.
\end{abstract}


\section{Introduction}
\label{Introduction}
One of the main goals of the study of galaxy formation and evolution is to understand the star formation history of the Universe.  A key advance in this area was the discovery of the cosmic far-infrared extragalactic background light (EBL) by the \emph{COBE} satellite \citep{Puget96,Fixsen98} with an energy density similar to that of the UV/optical EBL, implying that a significant amount of star formation over the history of the Universe has been obscured and its light reprocessed by dust. Following this, the population of galaxies now generally referred to as sub-millimetre galaxies (SMGs) was first revealed using the Sub-millimetre Common User Bolometer Array (SCUBA) on the James Clerk Maxwell Telescope \citep[JCMT, e.g.][]{Smail97,Hughes98}.  SMGs are relatively bright in sub-millimetre bands (the first surveys focussed on galaxies with $S_{850\mu\rm m}>5$ mJy) and some studies have now shown that the bulk of the EBL at 850$\mu \rm m$ can be resolved by the $S_{850\mu\rm m}>0.1$ mJy galaxy population \citep[e.g.][]{Chen13}.  SMGs are generally believed to be massive, dust enshrouded galaxies with extreme infrared luminosities ($L_{\rm IR}\gtrsim 10^{12} \rm{L}_{\odot}$) implying prodigious star formation rates (SFRs, $10^{2}$-$10^{3}$ $\rm{M}_{\odot}$yr$^{-1}$), though this is heavily dependent on the assumed stellar initial mass function \citep[IMF, e.g.][]{Blain02,Casey14}.

One difficulty for sub-millimetre observations is the coarse angular resolution ($\sim 20''$ FWHM) of ground-based single-dish telescopes used for many blank-field surveys.  Recently, follow-up surveys  performed with greater angular resolution ($\sim 1.5''$ FWHM) interferometers (e.g. Atacama Large Millimetre Array - ALMA, Plateau de Bure Interferometer - PdBI, Sub-Millimetre Array - SMA)  targeted at single-dish detected sources have indicated that the resolution of single-dish telescopes had in some cases blended the sub-mm emission of multiple galaxies into one single-dish source \citep[e.g.][]{Wang11,Smolcic12,Hodge13}.  \cite{Karim13} showed the effect this blending has on the observed sub-mm number counts, with the single-dish counts derived from the Large APEX (Atacama Pathfinder EXperiment) BOlometer CAmera (LABOCA) Extended \emph{Chandra} Deep Field-South (ECDFS) Sub-millimetre Survey \citep[LESS,][]{Weiss09} exhibiting a significant enhancement at the bright end relative to counts derived from the ALMA follow-up (ALESS).

A related observational difficulty concerning SMGs is determining robust multi-wavelength counterparts for single-dish sources.  This is in part due to the single-dish resolution spreading the sub-mm emission over a large solid angle making it difficult to pinpoint the precise origin to an accuracy of greater than $\pm 2''$.  This process is also compounded by the faintness of SMGs at other wavelengths. Sub-mm bands are subject to a negative $K$-correction, which results in the sub-mm flux of an SMG being roughly constant over a large range of redshifts $z\sim 1-10$ \citep[e.g.][]{Blain02}.  This negative $K$-correction is caused by the spectral energy distribution (SED) of a galaxy being a decreasing power law with wavelength where it is sampled by observer-frame sub-mm bands. As the SED is shifted to higher redshifts it is sampled at a shorter rest-frame wavelength, where it is intrinsically brighter.  This largely cancels out the effect of dimming due to the increasing luminosity distance.  When observed at other wavelengths e.g. radio, galaxies are subject to a positive $K$-correction and so become fainter with increasing redshift.  This is problematic as radio emission has often been used to aid in measuring the position of the sub-mm source, as the star formation that powers the dust emission in the sub-mm also produces radio emission from synchrotron electrons produced by the associated supernovae explosions. This radio selection technique thus biases the counterpart identification towards lower redshift \cite[e.g.][]{Chapman05}.  Typically, radio-identification yields robust counterparts for $\sim 60$\% of an SMG sample \citep[e.g.][]{Biggs11}.  Sub-mm interferometers have greatly improved the situation, providing positional accuracies of up to $\sim 0.2''$, free from any biases introduced by selection criteria at wavelengths other than the sub-mm.  Once multi-wavelength counterparts have been identified, photometric redshifts are derived through fitting an SED to the available photometry, allowing redshift to vary as a free parameter \cite[e.g.][]{Smolcic12}.  Whilst observationally inexpensive and thus desirable for large SMG surveys, the errors from photometric redshifts are often significant, and samples are again biased by requiring detection in photometric bands.

Compounding these difficulties is the fact that, with the exception of the South Pole Telescope (SPT) survey presented in \cite{Vieira10}\footnote{These authors surveyed 87~deg$^2$ at 1.4 (2) mm to a depth of 11~(4.4)~mJy with a 63$''$~(69$''$)~FWHM beam. Due to the flux limits and wavelength of this survey, the millimetre detections are mostly gravitationally lensed sources \citep{Vieira13}.}, ground-based sub-mm surveys have to date been pencil beams ($<0.7$~deg$^2$) leaving interpretation of the observed results subject to field-to-field variations.  In particular, \cite{Michalowksi12} found evidence that photometric redshift distributions of radio-identified counterparts of $1100$ and $850~\mu$m selected SMGs in the two non-contiguous SCUBA Half-Degree Extragalactic Survey (SHADES) fields are inconsistent with being drawn from the same parent distribution.  This suggests that the SMGs are tracing different large scale structures in the two fields.  Larger surveys have been undertaken at $250$, $350$ and $500$ $\mu$m from space using the Spectral and Photometric Imagine REceiver \citep[SPIRE,][]{Griffin10}  instrument on board the \emph{Herschel} Space Observatory \citep{Pilbratt10}.  These are also affected by coarse angular resolution; the SPIRE beam has a FWHM of $\sim18''$, $25''$ and $37''$ at 250, 350 and 500 $\mu$m respectively.  However, number counts at these wavelengths have been derived from SPIRE maps through stacking analysis \citep{Bethermin12} using the positions and flux densities of sources detected at $24~\mu$m as a prior.  

Historically, hierarchical galaxy formation models have struggled to reproduce the high number density of the SMG population at high redshifts \citep[e.g.][]{Blain99,DevriendtGuiderdoni00,Granato00}. However \cite{Baugh05} presented a version of the Durham semi-analytic model (SAM), \galform, which could successfully reproduce the observed number counts and redshift distribution of SMGs, along with the present day luminosity function.  In order to do so, it was found necessary to significantly increase the importance of high-redshift starbursts in the model relative to previous versions of \galform; this was primarily achieved through introducing a top-heavy stellar initial mass function (IMF) for galaxies undergoing a (merger induced) starburst.  Recently, \cite{Hayward13a} introduced a hybrid model which combined the results from idealized hydrodynamical simulations of isolated discs/mergers with various empirical cosmological relations and showed reasonable agreement with the $850$ $\mu\mathrm{m}$ number counts and redshift distribution utilising a solar neighbourhood IMF.  However, this model is limited in terms of the range of predictions it can make due to its semi-empirical nature.  A similar model was presented in \cite{Hayward13b} which included a treatment of blending by single dish telescopes, showing that the sub-mm emission from both physically associated and unassociated SMGs contribute significantly to the single-dish number counts.  This model underpredicts the observed single-dish number counts at $S_{850\mu\mathrm{m}}>5$ mJy, possibly due to the exclusion of starburst galaxies.  The Hayward et al. models build on earlier work presented in \cite{Hayward11} and \cite{Hayward12} which were novel in discussing theoretically the effects of the single-dish beam on the observed SMG population.

Here we investigate the effect of both the angular resolution of single-dish telescopes and field-to-field variations on observations of the SMG population.  We utilise $50$ randomly orientated lightcones calculated from an updated version of \galform (Lacey et al. 2014, in preparation, hereafter L14) to create mock sub-mm surveys taking into account the effects of the single-dish beam.  This paper is structured as follows: in Section \ref{sec:Model} we introduce the theoretical model we use for this analysis and our method for creating our $850$ $\mu$m  mock sub-mm surveys.  In Section \ref{sec:results} we present our main results concerning the effects of the single-dish beam and field-to-field variance.  In Section \ref{sec:ALESS} we make a detailed comparison of the predictions of our model with the ALESS survey and in Section \ref{sec:multi_lam} we present our predicted single-dish number counts at $450$ and $1100$ $\mu$m.   We summarise our findings and conclude in Section \ref{sec:Summary}.       

\section{The Theoretical Model}
\label{sec:Model}
In this section we present the model used in this work.  We couple a state-of-the-art semi-analytic galaxy formation model run in a Millennium-class \citep{Springel05} $N$-body simulation using the WMAP7 cosmology \citep{Komatsu11}\footnote{$\Omega_{0}=0.272$, $\Lambda_{0}=0.728$, $h=0.704$, $\Omega_{\rm b}=0.0455$, $\sigma_{8}=0.81$, $n_{s}=0.967$.  This is the simulation referred to as MS-W7 in \cite{Guo13} and \cite{vgp14}; and as MW7 in \cite{Jenkins13}.  It is available on the Millennium database at: \url{http://www.mpa-garching.mpg.de/millennium}.}, with a simple model for the re-processing of stellar radiation by dust (in which the dust temperature is calculated self-consistently). A sophisticated lightcone treatment is implemented for creating mock catalogues of the simulated galaxies \citep{Merson13}.  We also describe our method for creating sub-mm maps from these mock catalogues, which include the effects of the single-dish beam size and instrumental noise, from which we extract sub-mm sources in a way that is consistent with what is done in observational studies. 

\subsection{GALFORM}
The Durham SAM, \galform, was first introduced in \cite{Cole00}.  Galaxy formation is modelled \emph{ab initio}, beginning with a specified cosmology and a linear power spectrum of density fluctuations and ending with predicted galaxy properties at a range of redshifts.  Galaxies are assumed to form within dark matter halos, with their subsequent evolution controlled by the merging history of the halo.  These halo merger histories can be calculated using a Monte Carlo technique following extended Press-Schechter formalism \citep{PCH08}, or (as is the case in this work) extracted directly from $N$-body dark matter only simulations \citep[e.g.][]{Helly03,Jiang14}.  Baryonic physics is modelled using a set of continuity equations that track the exchange of baryons between stellar, cold disc gas and hot halo gas components.  The main physical processes that are modelled include: (i) hierarchical assembly of dark matter halos; (ii) shock heating and virialization of gas in halo potential wells; (iii) radiative cooling and collapse of gas onto galactic discs; (iv) star formation from cold gas; (v) heating and expulsion of gas through feedback processes; (vi) chemical evolution of gas and stars; (vii) mergers of galaxies within halos due to dynamical friction; (viii) evolution of stellar populations using stellar population synthesis (SPS) models; and (ix) the extinction and reprocessing of stellar radiation due to dust.  As with other SAMs, the simplified nature of the equations that are used to characterise these complex and in some cases poorly understood physical processes introduce a number of parameters into the model. These parameters are constrained using a combination of simulation results and observational data, reducing enormously the available parameter space.  In particular, the strategy of \cite{Cole00} is that for a galaxy formation model to be deemed successful it must reproduce the present day ($z=0$) luminosity function in optical and near infra-red bands.  For a more detailed overview of SAMs see the reviews by \cite{Baugh06} and \cite{Benson10}.  

Several \galform models have appeared in the literature that adopt different values for the model parameters and in some cases include different physical processes.  For this work we adopt the model presented in L14  as it can reproduce a range of observational data (including $z=0$ luminosity functions in $b_{\rm J}$ and $K$-bands, see L14 for more details) and because it combines a number of important physical processes from previous \galform models.  These include the effects of AGN feedback inhibiting gas cooling in massive haloes \citep{Bower06}, and a star formation law for galaxy discs \citep{Lagos11} based on an empirical relationship between the star formation rate and molecular-phase gas density \citep{BlitzRosolowsky06}.  For the purposes of  reproducing a number density of sub-mm galaxies appropriate for this study, a top-heavy IMF is implemented for starbursts, as in \cite{Baugh05}. However, in L14 a much less extreme slope is used compared to that invoked by Baugh et al\footnote{The slope of the IMF, $x$, in $dN(m)/d\ln m = m^{-x}$, has a value of $x=1$ in L14 whereas a value of $x=0$ was used in \cite{Baugh05}.}.  The top-heavy IMF enhances the sub-mm luminosity of a starburst galaxy through a combination of an enhanced number of massive stars which increases the unattenuated UV luminosity of the galaxy, and a greater number of supernovae events which increases the metal content and hence dust mass available to absorb and re-emit the stellar radiation at sub-mm wavelengths.  A significant difference between \cite{Baugh05} and L14 is that in Baugh et al. the starburst population was induced by galaxy mergers, whilst in L14 starbursts are primarily caused by disc instabilities.  These instabilities use the same stability criterion for self-gravitating discs presented in \cite{Mo98} and \cite{Cole00}.  They were included in \cite{Bower06}, but were not considered in \cite{Baugh05}.  As with other \galform models, a standard \cite{Kennicutt83} IMF is adopted in L14 for quiescently star forming discs.  

The model presented in L14 is designed to populate a Millennium-class dark matter only $N$-body simulation using a WMAP7 cosmology with a minimum halo mass of $1.9\times10^{10}$~$h^{-1}$~M$_{\odot}$.  This work uses $50$ output snapshots from the model in the redshift range $z=0{-}8.5$, we use this large redshift range so that our simulated SMG population is complete.    

\subsection{The Dust Model}
\label{sec:dust_model}
In order for the sub-mm flux of galaxies to be predicted, a model is required to calculate the amount of stellar radiation absorbed by dust and the resulting SED of the dust emission.  Here we use a model motivated by the radiative transfer code \grasil \citep{Silva98}. \grasil calculates the heating and cooling of dust grains of varying sizes and compositions at different locations within each galaxy, effectively obtaining the dust temperature $T_{\rm d}$ at each position.  \grasil has been coupled with \galform in previous works \citep[e.g.][]{Granato00,Baugh05,Swinbank08}. However, due to the computational expense of running \grasil for the number of \galform galaxies generated in the simulation volume used in this work, we instead use a model which retains some of the key assumptions of \grasil but with a significantly simplified calculation.  Despite the simplifications made, this model can accurately reproduce the predictions of \grasil for rest-frame wavelengths $\lambda_{\rm rest}>70$ $\mu\mathrm{m}$. We are therefore confident in its application to the wavelengths under investigation here. We briefly describe our dust model in the following section. However, for a more detailed explanation we refer the reader to the appendix of L14.

We adopt the \grasil assumptions regarding the geometry of the stars and dust.  Stars are distributed throughout two components (i) a spherical bulge with an $r^{1/4}$ density profile, and (ii) a flattened component which is either a quiescent disc or a starburst component, with exponential radial and vertical density profiles.  Young stars and dust are assumed to be in the flattened component only.  A two phase dust medium is also adopted, as in \grasil. Dust and gas exist in either dense molecular clouds, modelled as uniform density spheres of fixed mass ($10^6$ $\rm{M}_{\odot}$) and radius (16 pc), or a diffuse inter-cloud medium.  Stars are assumed to form inside the molecular clouds and gradually escape into the diffuse dust on a timescale of a few Myrs, parametrised as $t_{\rm esc}$ in the model.  The dust emission is first obtained by calculating the energy from stellar radiation absorbed in each dust component.  Assuming thermal equilibrium, this is then equated to the energy emitted by the respective dust component, such that the luminosity per unit wavelength emitted by a mass $M_{\rm d}$ of dust is given by
\begin{equation}
L_{\lambda}^{\rm{dust}}=4\pi\kappa_{\rm d}(\lambda)B_{\lambda}(T_{\rm d})M_{\rm d},
\end{equation} 
where $\kappa_{\rm d}(\lambda)$ is the absorption cross-section per unit mass and $B_{\lambda}(T_{\rm d})$ is the Planck blackbody function.  Crucially this means that the dust temperature of each component is not a free parameter but is calculated self-consistently, based on global energy balance arguments.  An important simplifying assumption here is that we assume only two dust temperatures, one for the molecular clouds and one for the diffuse medium.  The dust mass, $M_{\rm d}$, is proportional to the metallicity times the cold gas mass, normalised to give the local inter-stellar medium dust-to-gas ratio for solar metallicity.  For calculating dust emission, the dust absorption cross-section per unit mass of metals in the gas phase is approximated as follows:
\begin{equation}
\label{eq:kappa_d}
\kappa_{\rm d}(\lambda)=
\begin{cases}
\kappa_{1}\left(\frac{\lambda}{\lambda_{1}}\right)^{-2}&\lambda<\lambda_{\rm b}\\
\kappa_{1}\left(\frac{\lambda_{\rm b}}{\lambda_{1}}\right)^{-2}\left(\frac{\lambda}{\lambda_{\rm b}}\right)^{-\beta_{\rm b}}&\lambda>\lambda_{\rm b}\mathrm{.}\\
\end{cases}
\end{equation}
With $\kappa_{1}=140$ cm$^2$g$^{-1}$ at the reference wavelength $\lambda_{1}=30$ $\mu$m \citep[e.g.][]{DraineLee84}.  The power-law break is introduced at $\lambda_{\rm b}=100$ $\mu$m for starburst galaxies \emph{only}, with $\beta_{\rm b}=1.5$. For quiescently star forming galaxies we assume an unbroken power law, equivalent to $\lambda_{\rm b}\rightarrow\infty$.

The sub-mm number counts can be calculated by first constructing luminosity functions $dn/d\ln L_{\nu}$ at a given output redshift using $L_{\nu}$ calculated by the dust model.  The binning in luminosity is chosen so that we have fully resolved the bright end, to which the derived number counts are sensitive. The number counts and redshift distribution can then be calculated using
\begin{equation}
	\frac{d^2 N}{d\ln S_{\nu}dzd\Omega} = \left\langle\frac{dn}{d \ln L_{\nu}}\right\rangle\frac{dV}{dzd\Omega},
	\label{eq:ncts}
\end{equation}
where the comoving volume element $dV/dz=(c/H(z))r^2(z)$, $r(z)$ is the comoving radial distance to redshift $z$, and the brackets $\langle...\rangle$ represent a volume-averaging utilising the whole $N$-body simulation volume (500 $h^{-1}$Mpc)$^3$.

\subsection{Creating mock surveys}
\label{sec:create_mock_surveys}
In order to create mock catalogues of our sub-mm galaxies we utilise the lightcone treatment described in \cite{Merson13}.  Briefly,  as the initial simulation volume side-length ($L_{\rm{box}}=500$ $h^{-1}$Mpc) corresponds to the co-moving distance out to $z\sim 0.17$, the simulation is periodically replicated in order to fully cover the volume of a typical SMG survey, which extends to much higher redshift.  This replication could result in structures appearing to be repeated within the final lightcone, which could produce unwanted projection-effect artefacts if their angular separation on the `mock sky' is small \citep{Blaizot05}.  As our fields are small in solid angle (0.5 deg$^2$) and our box size is large, we expect this effect to be of negligible consequence and note that we have seen no evidence of projection-effect artefacts in our mock sub-mm maps.  Once the simulation volume has been replicated, a geometry is determined by specifying an observer location and lightcone orientation. An angular cut defined by the desired solid angle of our survey is then applied, such that the mock survey area resembles a sector of a sphere.  The redshift of a galaxy in the lightcone is calculated by first determining the redshift ($z$) at which its host dark matter halo enters the observer's  past lightcone.  The positions of galaxies are then interpolated from the simulation output snapshots ($z_{i},z_{i+1}$, where $z_{i+1}<z<z_{i}$) such that the real-space correlation function of galaxies is preserved. A linear $K$-correction interpolation is applied to the luminosity of the galaxy to account for the shift in $\lambda_{\rm rest}=\lambda_{\rm obs}/(1+z)$ for a given $\lambda_{\rm obs}$, based on its interpolated redshift.

\begin{figure}
\includegraphics[width=\columnwidth]{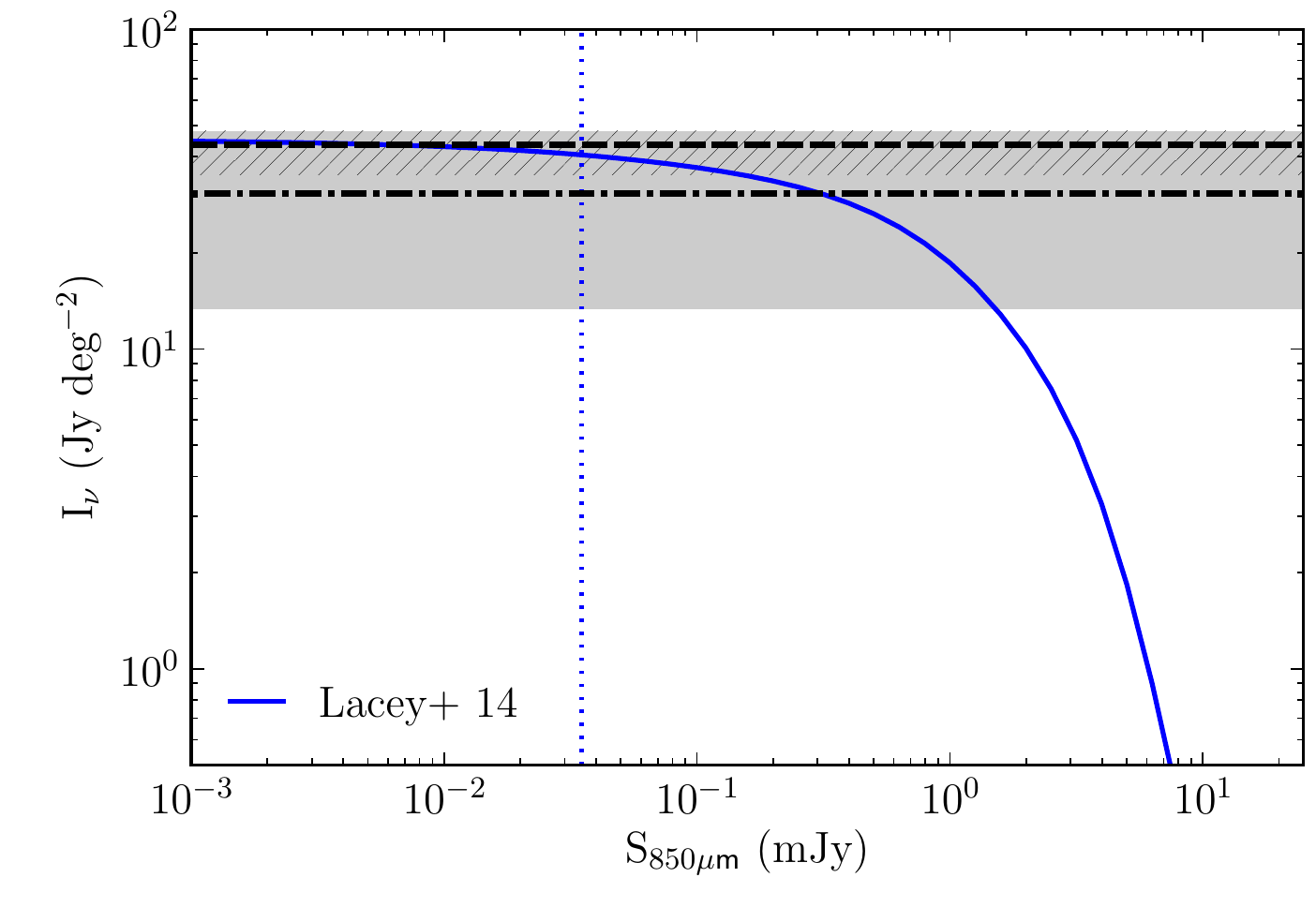}
\caption{Predicted cumulative extragalactic background light as a function of flux at 850 $\mu$m (blue line).  The horizontal dashed line \citep{Fixsen98} and dash-dotted line \citep{Puget96} show the background light as measured by the \emph{COBE} satellite.  The shaded \citep{Puget96} and hatched \citep{Fixsen98} regions indicate the respective errors on the two measurements.  The vertical dotted line indicates the flux limit above which $90\%$ of the total predicted EBL is resolved.}
\label{fig:EBL}
\end{figure}

\begin{figure}
\includegraphics[width=\columnwidth]{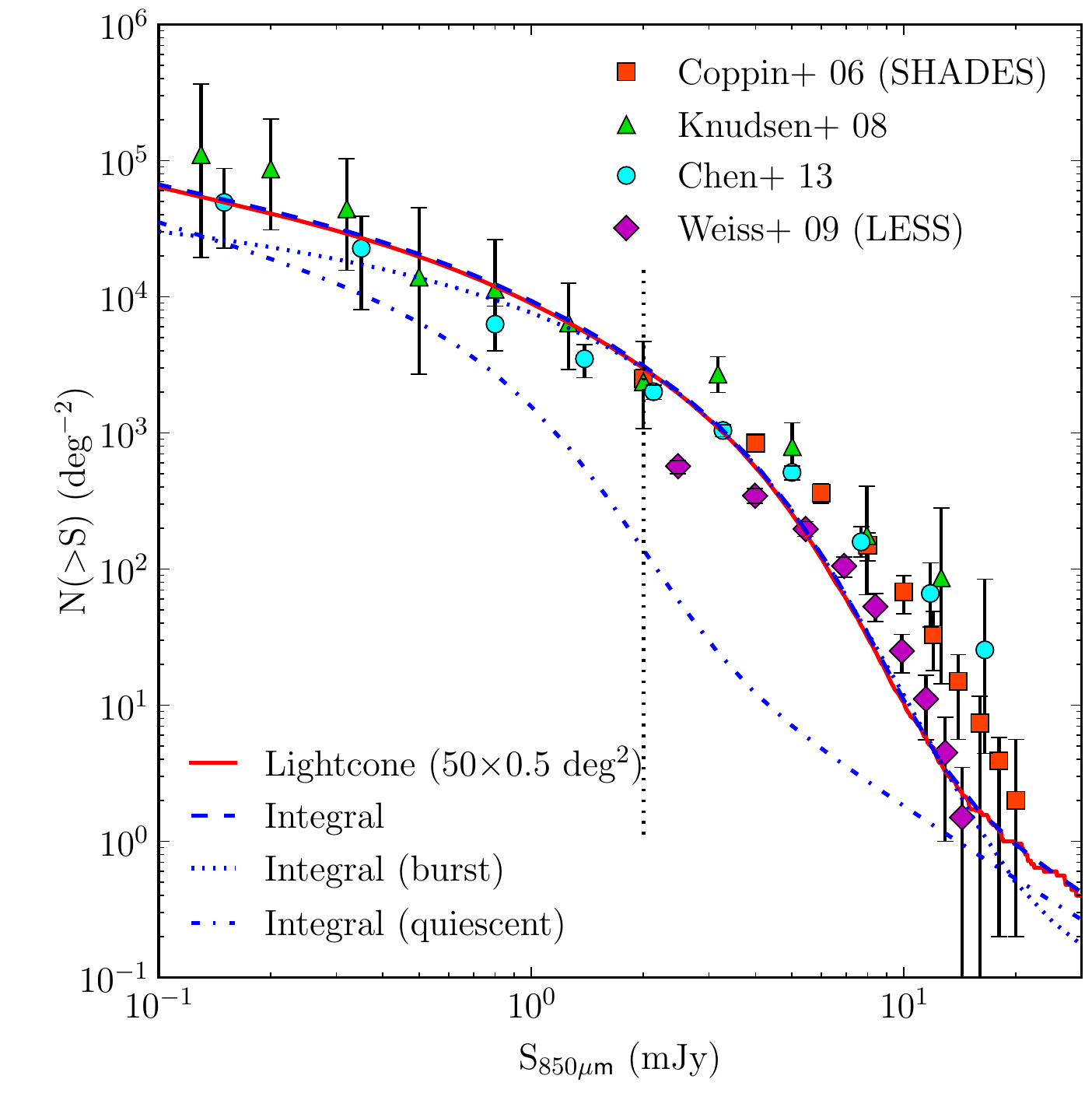}
\caption{Predicted cumulative number counts at 850 $\mu$m.  Predictions from the lightcone catalogues (red line) and from integrating the luminosity function of the model (dashed blue line) are in excellent agreement.  The dotted and dash-dotted blue lines show the contribution to the number counts from starburst and quiescent galaxies respectively.  We compare the model predictions to single-dish observational data from Coppin at al. (\citeyear{Coppin06}; orange squares), Knudsen et al. (\citeyear{Knudsen08}; green triangles), Wei{\ss} et al. (\citeyear{Weiss09}; magenta diamonds) and Chen at al. (\citeyear{Chen13}; cyan circles).  The vertical dotted line shows the approximate confusion limit ($\sim2$ mJy) of single-dish blank field surveys. Observational data fainter that this limit are derived from cluster-lensed surveys (see Section \ref{sec:nctsbeam} for further discussion).}
\label{fig:ncts_int_lc}
\end{figure}

To create the 850 $\mu$m mock catalogues we apply a further selection criterion so that our galaxies have $S_{850\mu\rm m}>0.035$ mJy. This is the limit brighter than which we recover $\sim 90\%$ of the 850 $\mu\rm m$ EBL, as predicted by our model (Fig. \ref{fig:EBL}).     
We have checked that our simulated SMG population is not affected by incompleteness at this low flux limit, due to the finite halo mass resolution of the $N$-body simulation. To allow us to test field-to-field variance we generate $50~\times~0.5$ deg$^2$ lightcone surveys\footnote{In practise our surveys are $0.55$ deg$^2$.  This allows for galaxies outside the $0.5$ deg$^2$ area to contribute to sources detected inside this area after convolution with the single-dish beam.} with random observer positions and lines of sight.  In Fig. \ref{fig:ncts_int_lc} we show that the lightcone accurately reproduces the SMG number counts of our model.  We also show in Fig. \ref{fig:ncts_int_lc} the predicted $850~\mu$m number counts from starburst (dotted line) and quiescent (dash-dotted line) galaxies in the model.  Starburst galaxies dominate the number counts in the range $\sim0.2{-}20$ mJy.  Turning off merger-triggered starbursts in this model has a negligible effect on the predicted number counts (L14), from this we have inferred that these bursts are predominately triggered by disc instabilities.  
\subsection{Creating sub-mm maps}
\label{sec:create_submm_maps}
\begin{figure*}
\centering
\includegraphics[width=0.8\linewidth]{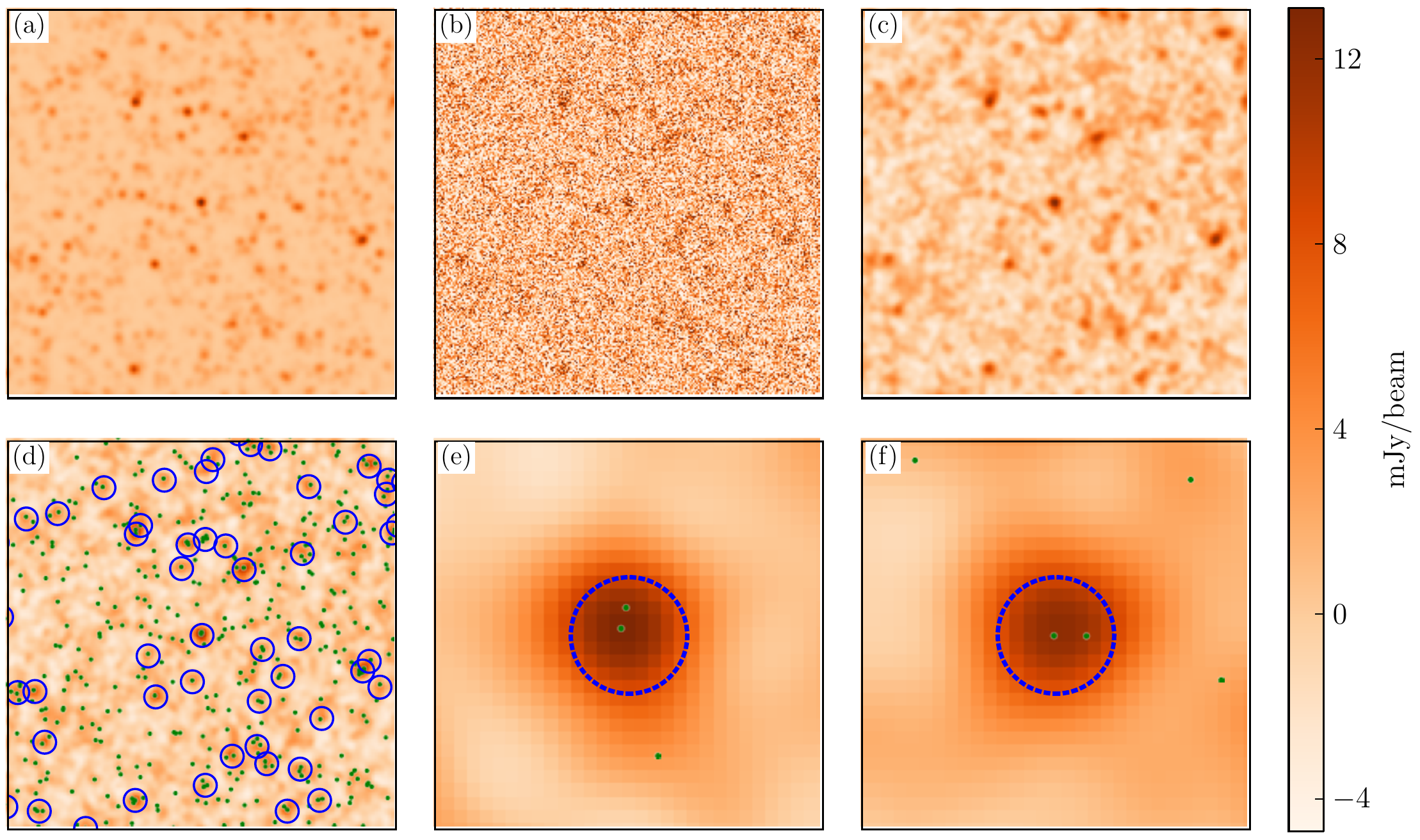}
\caption{Panels illustrating the mock map creation process at 850 $\mu$m.  Panels (a)-(d) are $0.2\times 0.2$ deg$^2$ and are centred on a $13.1$ mJy source. (a) Astrophysical map including the effect of the telescope beam. (b) Astrophysical plus Gaussian white noise map, constrained to have zero mean. (c) Matched-filtered map. (d) Matched-filtered map with $S_{850\mu\rm m}>4$ mJy single-dish sources (blue circles centred on the source position) and $S_{850\mu \rm m}>1$ mJy galaxies (green dots) overlaid. (e) As for (d) but for a $0.5'\times0.5'$ area, centered on the same 13.1~mJy source.  The 2 galaxies within the $9''$ radius (blue dotted circle, $\sim$ ALMA primary beam) of the source have fluxes of $1.2$ and $11.2$ mJy and redshifts of $1.0$ and $2.0$ respectively.  (f) as for (e) but centred on a $12.2$ mJy source.  In this case the 2 galaxies within the central $9''$ radius have fluxes of $6.1$ and $6.4$ mJy and redshifts of $2.0$ and $3.2$ respectively.}   
\label{fig:thumbs}
\end{figure*}

Here we describe the creation of mock sub-mm maps from our lightcone catalogues.  First, we create an image by assigning the $850$ $\mu$m flux of a galaxy to the pixel in which it is located, using a pixel size much smaller than the single-dish beam.  This image is then convolved with a point spread function (PSF), modelled as a 2D Gaussian with a $15''$ FWHM ($\sim$SCUBA2/JCMT), and then re-binned into a coarser image with $2''\times2''$ pixels, to match observational pixel sizes.  The resulting image is then scaled so that it is in units of mJy/beam.  We refer to the output of this process as the astrophysical map (see Fig \ref{fig:thumbs}a).

In order to model the noise properties of observational maps we add `instrumental' Gaussian white noise to the astrophysical map. We tune the standard deviation of this noise such that after it has been matched-filtered (described below) the output is a noise map with $\sigma_{\rm rms} \sim 1$ mJy/beam, comparable to jackknifed noise maps in $850$ $\mu$m blank-field observational surveys \citep[e.g.][]{Coppin06,Weiss09,Chen13}. 
 
It is a well known result in astronomy that the best way to find point-sources in the presence of noise is to convolve with the PSF \citep{Stetson87}. However, this is only optimal if the noise is Gaussian, and does not take into account `confusion noise' from other point-sources.  \cite{Chapin11} show how one can optimise filtering for maps with significant confusion, through modelling this as a random (and thus un-clustered) superposition of point sources convolved with the PSF, normalised to the number counts inferred from $P(D)$ analysis of the maps.  The PSF is then divided by the power spectrum of this confusion noise realisation.  This results in a matched-filter with properties similar to a `Mexican-hat' kernel.  An equivalent method is implemented in \cite{Laurent05}.  Although our simulated maps contain a significant confusion background, for simplicity we do not implement such a method here, and have checked that the precise method of filtering does not significantly affect our source-extracted number counts. 

Prior to source extraction, we constrain our astrophysical plus Gaussian noise map to have a mean of zero (Fig. \ref{fig:thumbs}b) and convolve with a matched-filter $g(x)$, given by
\begin{equation}
g(x)=\mathcal{F}^{-1} \left\{ \frac{s^{*}(q)}{\int |s(q)|^2 d^{2}q} \right\},
\label{eq:filter}
\end{equation} 
where $\mathcal{F}^{-1}$ denotes an inverse Fourier transform, $s(q)$ is the Fourier transform of our PSF and the asterisk indicates complex conjugation. The denominator is the appropriate normalisation such that peak heights of PSF-shaped sources are preserved after filtering.  Up to this normalisation factor, the matched-filtering is equivalent to convolving with the PSF. Point sources are therefore effectively convolved with the PSF twice, once by the telescope and once by the matched-filter.  This gives our final matched-filtered map (Fig. \ref{fig:thumbs}c) a spatial resolution of $\sim 21.2''$ FWHM i.e. $\sqrt{2}\times 15''$.  

For real surveys, observational maps often have large scale filtering applied prior to the matched-filtering described above.  This is to remove large scale structure from the map, often an artefact of correlated noise of non-astrophysical origin.  This is implemented by convolving the map with a Gaussian broader than the PSF and then subtracting this off the original, rescaling such that the flux of point sources is conserved \citep[e.g.][]{Weiss09,Chen13}.  As our noise is Gaussian, any excess in the power spectrum of the map on large scales can only be attributed to our astrophysical clustering signal, so we choose not to implement any such high-pass filtering prior to our matched-filtering. 

\begin{figure}
\centering
\includegraphics[width=\columnwidth]{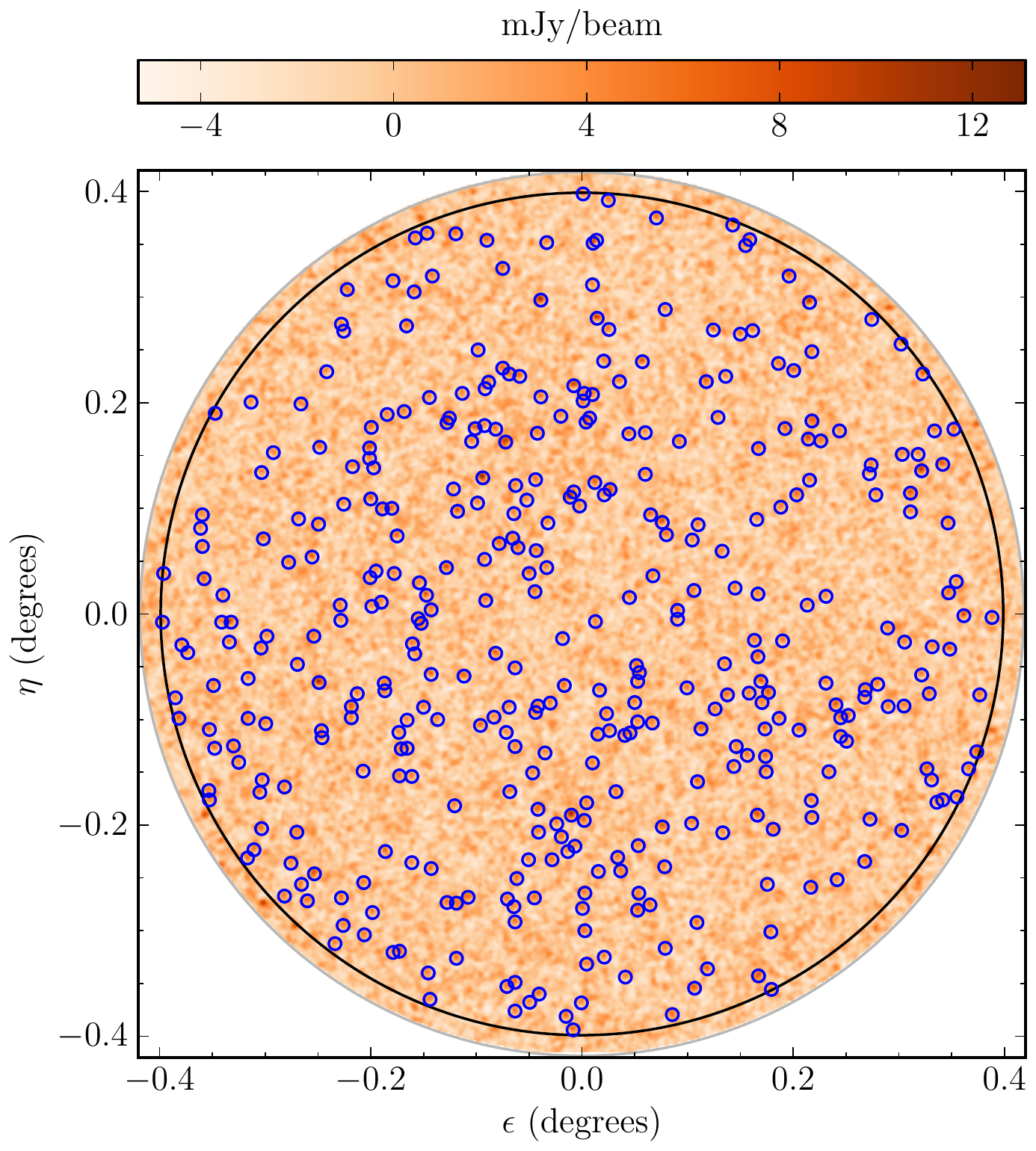}
\caption{A matched-filtered map.  Sources detected with $S_{850\mu\rm m}>4.5$ mJy by our source extraction algorithm are indicated by blue  circles.  The central $0.5$ deg$^2$ region, from which we extract our sources, is indicated by the black circle.}
\label{fig:Map}
\end{figure}

\begin{figure}
\centering
\includegraphics[width=\columnwidth]{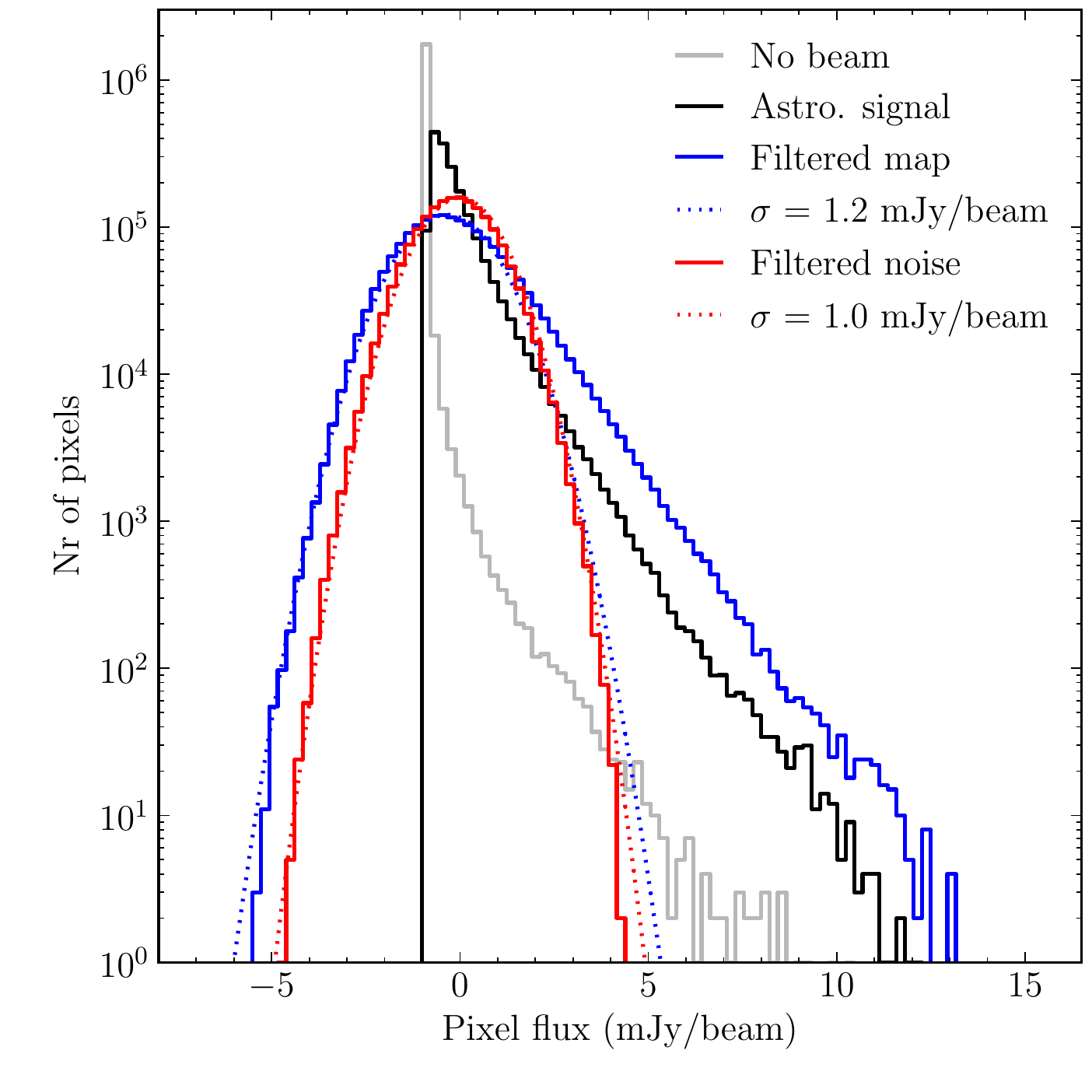}
\caption{Pixel flux histogram of the map shown in Fig. \ref{fig:Map}.  The grey and black lines are the map before and after convolution with the single-dish beam respectively, with the same zero point subtraction applied as to our final matched-filtered map (blue line).  The map is rescaled after convolution with the single-dish beam to convert to units of mJy/beam (grey to black), and during the matched-filtering due to the normalisation of the filter which conserves point source peaks (black to blue).    Dotted lines show Gaussian fits to the matched-filtered noise-only (red solid line) and  the negative tail of the final matched-filtered (blue solid line) map histograms respectively.}
\label{fig:pixelplot}
\end{figure}

An example of one of our matched-filtered maps is shown in Fig. \ref{fig:Map} and the associated pixel histogram in Fig. \ref{fig:pixelplot}.  The position of the peak of the pixel histogram is determined by the constraint that our maps have a zero mean after subtracting a uniform background.  We attribute the broadening of the Gaussian fits from $\sigma=1$ mJy/beam in our matched-filtered noise-only map to $\sigma=1.2$ mJy/beam in our final matched-filtered map to the realistic confusion background from unresolved sources in our maps. 

For the source extraction we first identify the peak (i.e. brightest) pixel in the map. For simplicity we record the source position and flux to be the centre and value of this peak pixel. We then subtract the matched-filtered PSF, scaled and centred on the value and position of the peak pixel, from our map.  This process is iterated down to an arbitrary threshold value of $S_{850\mu\rm m}=1$ mJy, resulting in our source-extracted catalogue.

\section{Results}
In this section we present our main results: in Section \ref{sec:nctsbeam} we show the effect the single-dish beam has on the predicted number counts through blending the sub-mm emission of  galaxies into a single source. In Section \ref{sec:multiplicity} we quantify the multiplicity of blended sub-mm sources, in Section \ref{sec:unassociated} we show that these blended galaxies are typically physically unassociated and in Section \ref{sec:dndz} we present the redshift distribution of SMGs in our model. 
\label{sec:results}
\subsection{Number counts}
\label{sec:nctsbeam}
\begin{figure}
\centering
\includegraphics[width=\columnwidth]{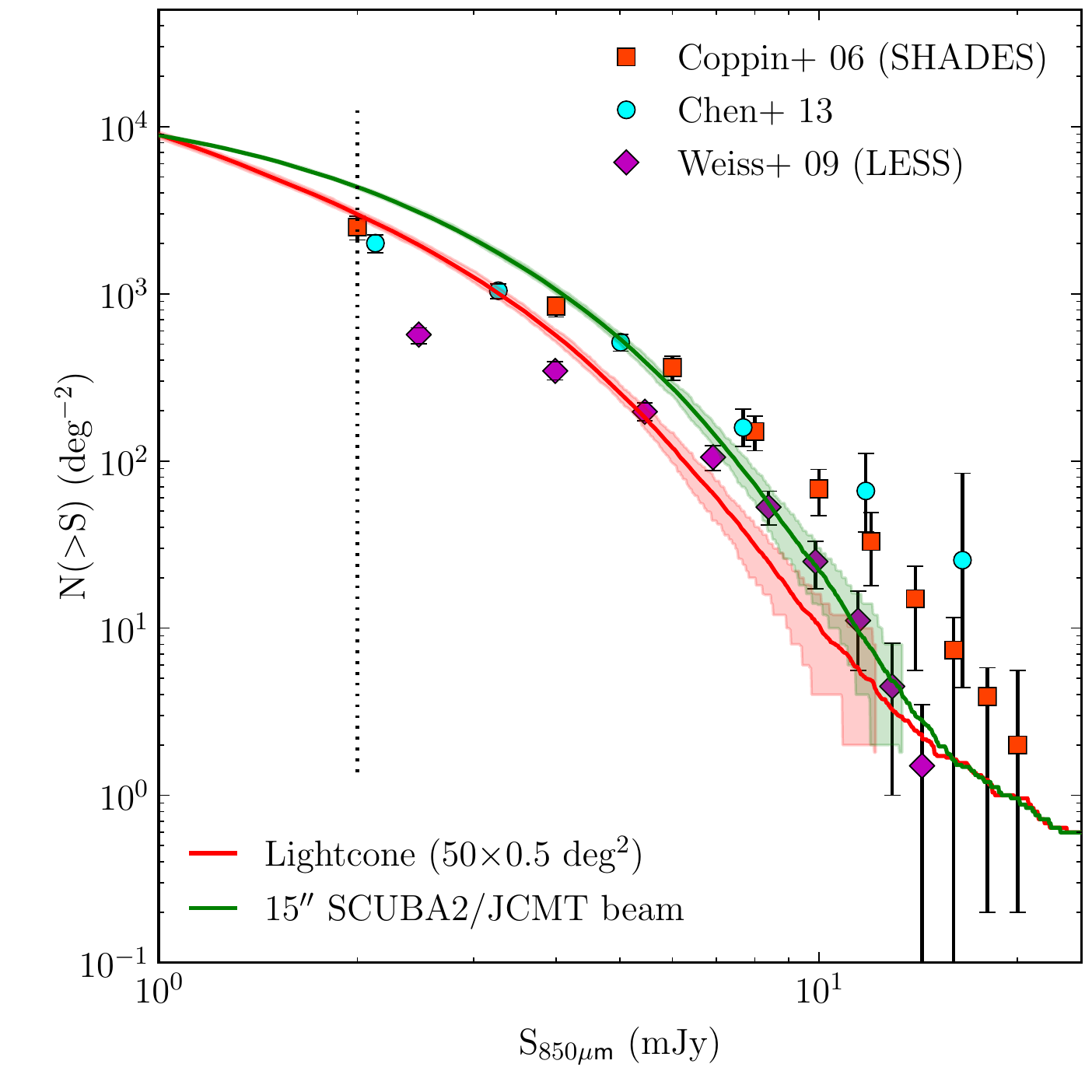}
\caption{The effect of single-dish beam size on cumulative 850 $\mu$m number counts.  The shaded regions show 10-90 percentiles of the distribution of the number counts from the $50$ individual fields, solid lines show counts from the combined 25 deg$^2$ field for the lightcone (red) and the $15''$ FWHM beam source extracted (green) catalogues.  The vertical dotted line at $S_{850 \mu \rm m}=2$ mJy indicates the approximate confusion limit of single-dish surveys. The $15''$ beam prediction is only to be compared at fluxes above this limit.  Single-dish blank field observational data is taken from Coppin et al. (\citeyear{Coppin06}; orange squares) Wei{\ss} et al. (\citeyear{Weiss09}; magenta diamonds) and Chen et al. (\citeyear{Chen13}; cyan circles).}
\label{fig:nctsbeam}
\end{figure}

The cumulative number counts derived from our lightcone and source-extracted catalogues are presented in Fig. \ref{fig:nctsbeam}.  The shaded regions, which show the 10-90 percentiles of the distribution of number counts from the individual fields, give an indication of the field-to-field variation we predict for fields of $0.5$ deg$^2$ area. This variation is comparable to or less than the quoted observational errors.  Quantitatively, we find a field-to-field variation in the source-extracted number counts of $0.07$ dex at 5 mJy and $0.34$ dex at 10 mJy.  A clear enhancement in the source-extracted number counts relative to those derived from our lightcone catalogues is evident at $S_{850\mu\rm m}\gtrsim 1$ mJy. We attribute this to the finite angular resolution of the beam blending together the flux from multiple galaxies with projected on-sky separations comparable to or less than the size of the beam.  Our source-extracted number counts show better agreement than our intrinsic lightcone counts with blank-field single-dish observational data above the confusion limit ($S_{\rm lim}\approx2$ mJy) of such surveys, which is indicated by the vertical dotted line in Fig \ref{fig:nctsbeam}.
 
Observational data fainter than this limit have been measured from gravitationally lensed cluster fields, where gravitational lensing due to a foreground galaxy cluster magnifies the survey area, typically by a factor of a few, but up to $\sim 20$.  The magnification increases the effective angular resolution of the beam, thus reducing the confusion limit of the survey and the instances of blended galaxies.  The lensing also boosts the flux of the SMGs. These effects allow cluster-lensed surveys to probe much fainter fluxes than blank-field surveys performed with the same telescope.  We show observational data in Fig. \ref{fig:ncts_int_lc} at $S_{850\mu\rm m}<2$ mJy for comparison with our lightcone catalogue number counts, with which they agree well.

Fig. \ref{fig:nctsbeam} shows that at $S_{850\mu\rm m}\gtrsim 5$ mJy our source-extracted counts agree best with the \cite{Weiss09} data, taken from ECDFS.  There is some discussion in the literature over whether this field is under-dense by a factor of $\sim 2$ (see Section 4.1 of \cite{Chen13} and references therein).  Whilst the field-to-field variation in our model can account for a factor of $\sim 2$ (at 10 mJy) it may be that our combined field source-extracted counts (and also those of Wei{\ss} et al.) are indeed underdense compared to number counts representative of the whole Universe.  

At $2\lesssim S_{850\mu \rm{m}}\lesssim5$ mJy our source-extracted number counts appear to follow a slightly steeper trend compared to the observed counts, this may be due to the underlying shape of our lightcone catalogue counts and the effect this has on our source-extracted counts.  We stress here that the L14 model was developed without regard to the precise effect the single-dish beam would have on the number counts. An extensive parameter search which shows the effect of varying certain parameters on the intrinsic number counts (and other predictions) of the model is presented in L14.  We do not consider any variants on the model here, but it is possible that once the effects of the single dish beam have been taken into account some variant models will match other observational data better, and show different trends over the flux range of interest.

The observed number counts at faint fluxes, above the confusion limit, may also be affected by completeness issues.  Whilst efforts are made to account for these in observational studies, they often rely on making assumptions about the number density and clustering of SMGs, so it is not clear that they are fully understood. 

\subsection{Multiplicity of single-dish sources}
\label{sec:multiplicity}
Given that multiple SMGs can be blended into a single source, in this section we quantify this multiplicity.  For each galaxy within a $4\sigma$ radius\footnote{We use the $\sigma$ of our match-filtered PSF i.e. $\sqrt{2}\times\rm{FWHM}/2\sqrt{2\ln{2}} \approx 9''$, and choose $4\sigma$ so that the search radius is large enough for our results in this section to have converged after our flux weighting scheme has been applied.} of a given $S_{850\mu\rm{m}} > 2$ mJy source, we determine a  flux contribution for that galaxy at the source position by modelling its flux distribution as  the matched-filtered PSF with a peak value equal to that galaxy's flux.  For example, a 5 mJy galaxy at a $\sim10.6''$ ($\sigma\times\sqrt{2\ln{2}}$) radial distance from a given source will contribute 2.5 mJy at the source position.   We  do this for all galaxies within the $4\sigma$ search radius and label the sum of these contributions  as the total galaxy flux of the source,  $S_{\rm gal\_tot}$.  The fraction each galaxy contributes towards this total is the galaxy's flux weight.  For each source we then interpolate the cumulative distribution of flux weights after sorting in order of decreasing flux weight, to determine how many galaxies are required to contribute a given percentage of the total.  

\begin{figure}
\centering
\includegraphics[width=\columnwidth]{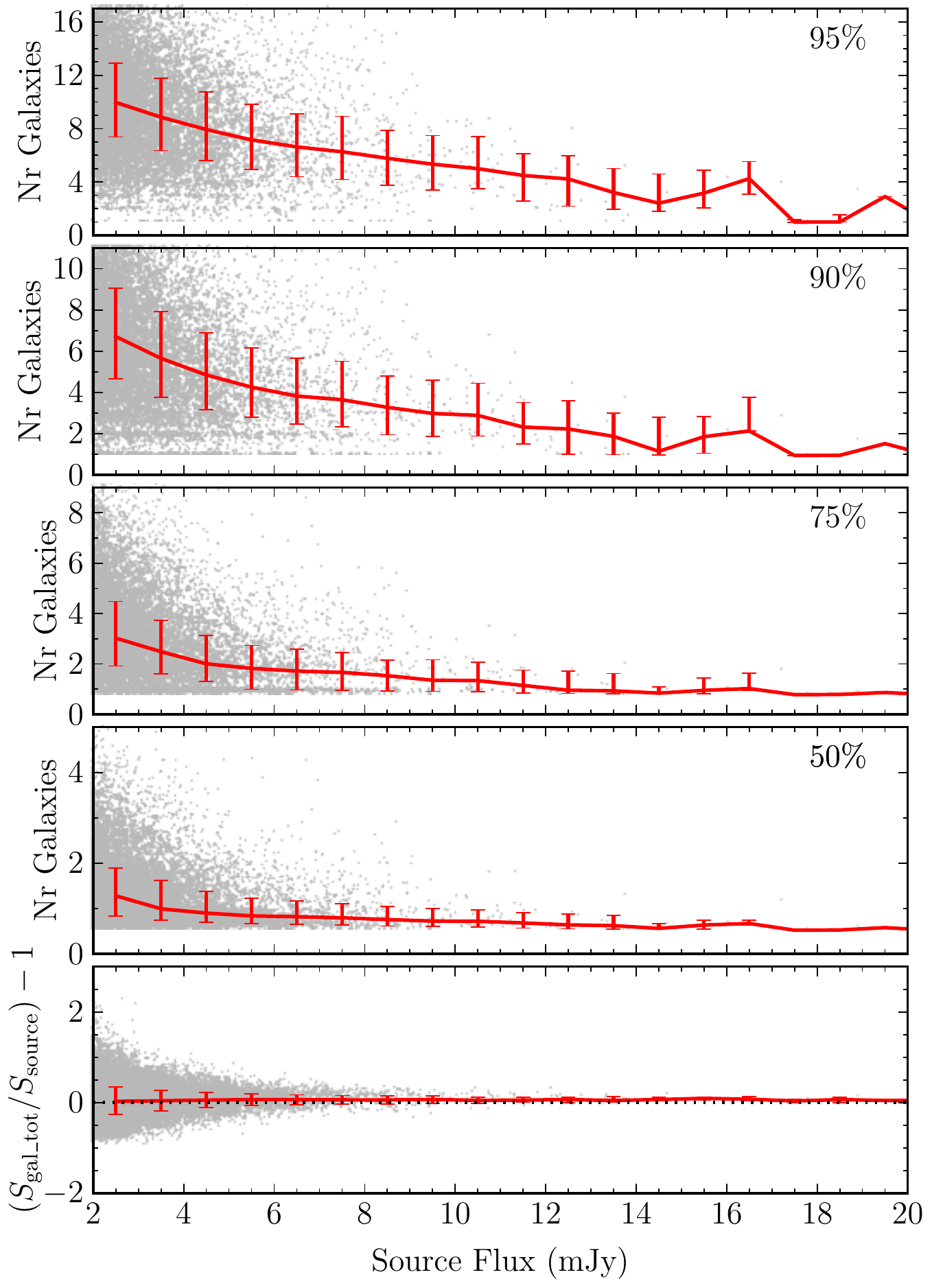}
\caption{\emph{Top $4$ panels}: Number of component galaxies contributing the percentage indicated in the panel of the total galaxy flux (see text) of a $S_{850 \mu m}>2$ mJy source.  \emph{Bottom panel}: Ratio of total galaxy flux to source flux.  Black dashed line is a reference line drawn at zero.  Solid red line shows median and errorbars indicate inter-quartile range for a 2mJy flux bin in all panels.  Grey dots show individual sources, for clarity only 10\% of the sources have been plotted.}
\label{fig:multiplicity}
\end{figure}

We plot this as a function of source-extracted flux, which includes the effect of instrumental noise and the subtraction of a uniform background, in the top $4$ panels of Fig \ref{fig:multiplicity}.  Typically, $90\%$ of the total galaxy flux of a $5$ mJy source is contributed by $\sim3{-}6$ galaxies and this multiplicity decreases slowly as source flux increases.  This decrease follows intuitively from the steep decrease in number density with increasing flux in the number counts.   

We note that this is not how source multiplicity is typically measured in observations. In Section \ref{sec:ALESS_ncts} we discuss the multiplicity of ALESS sources in a way more comparable to observations, where we have considered the flux limit and primary beam profile of ALMA, see also Table \ref{tab:ALMAtable}.  Observational interferometric studies which suggest that the multiplicity of single-dish sources may increase with increasing source flux \citep[e.g.][]{Hodge13} are likely to be affected by a combination of the flux limit of the interferometer, meaning high multiplicity faint sources are undetected, and small number statistics of bright sources.         

We also show, in the bottom panel of Fig. \ref{fig:multiplicity}, the ratio of the total galaxy flux to source flux. The consistency with zero indicates that our source-extracted number counts at 850 $\mu$m are not systematically biased.    This is due to the competing effects of subtracting a mean background in the map creation (which biases $S_{\rm source}$ low) and the introduction of Gaussian noise (which biases $S_{\rm source}$ high due to Eddington bias caused by the steeply declining nature of the number counts) effectively cancelling each other out in this case.  In Section \ref{sec:multi_lam} we find that our number counts at $450~\mu$m are strongly affected by Eddington bias, which we correct for in that case.          

\subsection{Physically unassociated galaxies}
\label{sec:unassociated}

\begin{figure}
\centering
\includegraphics[width=\columnwidth]{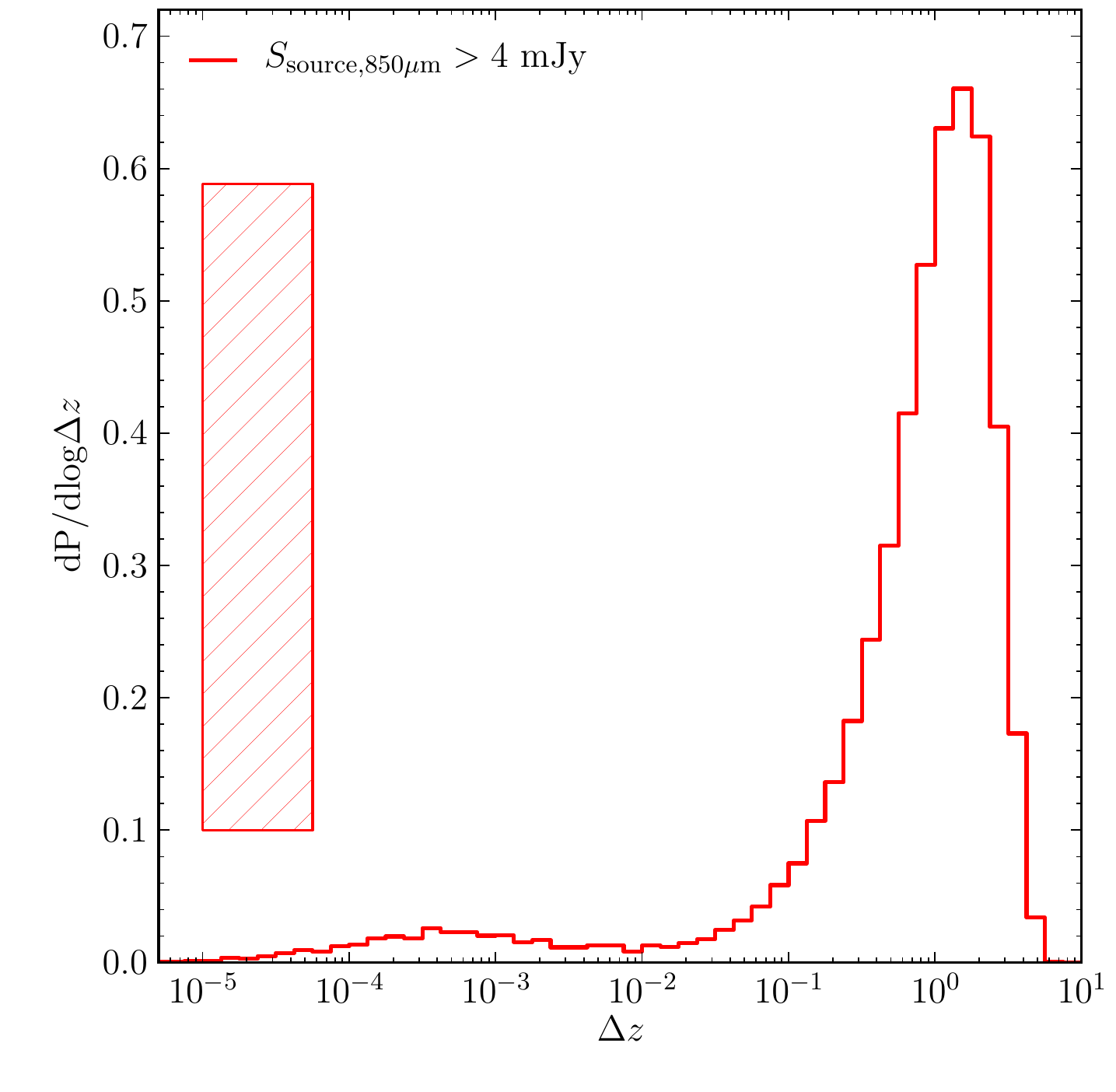}
\caption{Distribution of the logarithm of redshift separation (see text) of  $S_{850\mu\mathrm{m}}>4$ mJy single-dish sources.  The dominant peak at $\Delta z\approx1$ implies that the majority of the blended galaxies are physically unassociated.  The hatched region indicates the percentage ($\sim36\%$) of sources for which $\Delta z=0$ (see text in Section \ref{sec:unassociated}).}
\label{fig:associated}
\end{figure}

Given the multiplicity of our sources, we can further determine if the blended galaxies contributing to a source are physically associated, or if their blending has occurred due to a chance line of sight projection.  For each source we define a redshift separation, $\Delta z$, as the inter-quartile range of the cumulative distribution of the flux weights (calculated as described above),  where the galaxies have in this case first been sorted by ascending redshift.    The distribution of $\Delta z$ across our entire $S_{850\mu\mathrm{m}}>4$ mJy source population is shown in Fig. \ref{fig:associated}.  The dominant peak at $\Delta z\approx 1$ is similar to the distribution derived from a set of maps which  had galaxy positions randomised prior to convolution with the single-dish beam. This suggests that this peak is a result solely of a random sampling from the redshift distribution of our SMGs and thus that the majority of our sources are  composed of physically unassociated galaxies with a small on-sky separation due to chance line of sight projection.  This is unsurprising considering the large effective redshift range of sub-millimetre surveys, resulting from the negative $K$-corrections of SMGs.  We attribute the secondary peak at $\Delta z\sim 5\times 10^{-4}$ to clustering in our model, and defer a more thorough analysis of this to a future work.  We also show as the hatched region the area ($\sim36\%$) of sources for which $\Delta z=0$. These are sources for which a single galaxy spans the inter-quartile range of the cumulative distribution described above, this can occur when the flux weight of that galaxy is $>0.5$ and must occur when the flux weight of that galaxy is $>0.75$.   We understand that this is not how redshift separation would be defined observationally, and refer the reader to Section \ref{sec:ALESS} and Fig. \ref{fig:ALESS_delta_z} for another definition of $\Delta z$.  We note, however, that our conclusions in this section are not sensitive to the precise definition of $\Delta z$.

It is a feature of most current SAMs that any star formation enhancement caused by gravitational interactions of physically associated galaxies prior to a merger event is not included.  In principle this may affect our physically unassociated prediction, as in our model galaxy mergers would only become sub-mm bright post-merger, and would be classified as a single galaxy.  However, as merger induced starbursts have a negligible effect on our sub-mm number counts, which are composed of starbursts triggered by disc instabilities (L14), we are confident our physically unassociated conclusion is not affected by this feature.

We note that this conclusion is in contrast to predictions made by \cite{Hayward13b} who, in addition to physically unassociated blends, predict a more significant physically associated population than is presented here.  However, we believe our work has a number of significant advantages over that of \cite{Hayward13b} in that: (i) galaxy formation is modelled here \emph{ab initio} with a model that can also successfully reproduce galaxy luminosity functions at $z=0$; (ii) the treatment of blending presented here is more accurate through convolution with a beam, the inclusion of instrumental noise and matched-filtering prior to source-extraction, rather than a summation of sub-mm flux within some radius around a given SMG; and (iii) our $15''$ source-extracted number counts show better agreement with single-dish data for $S_{850\mu\rm m}\gtrsim 5$ mJy, this is probably in part due to the exclusion of starbursts from the Hayward et al. (2013b) model, though the effect including starbursts would have on the number counts in that model is not immediately clear.  

\subsection{Redshift distribution}
\label{sec:dndz}
As we have shown that sub-mm sources are composed of multiple galaxies at different redshifts, for this section we consider our lightcone catalogues \emph{only}.  

\begin{figure}
\centering
\includegraphics[width=\columnwidth]{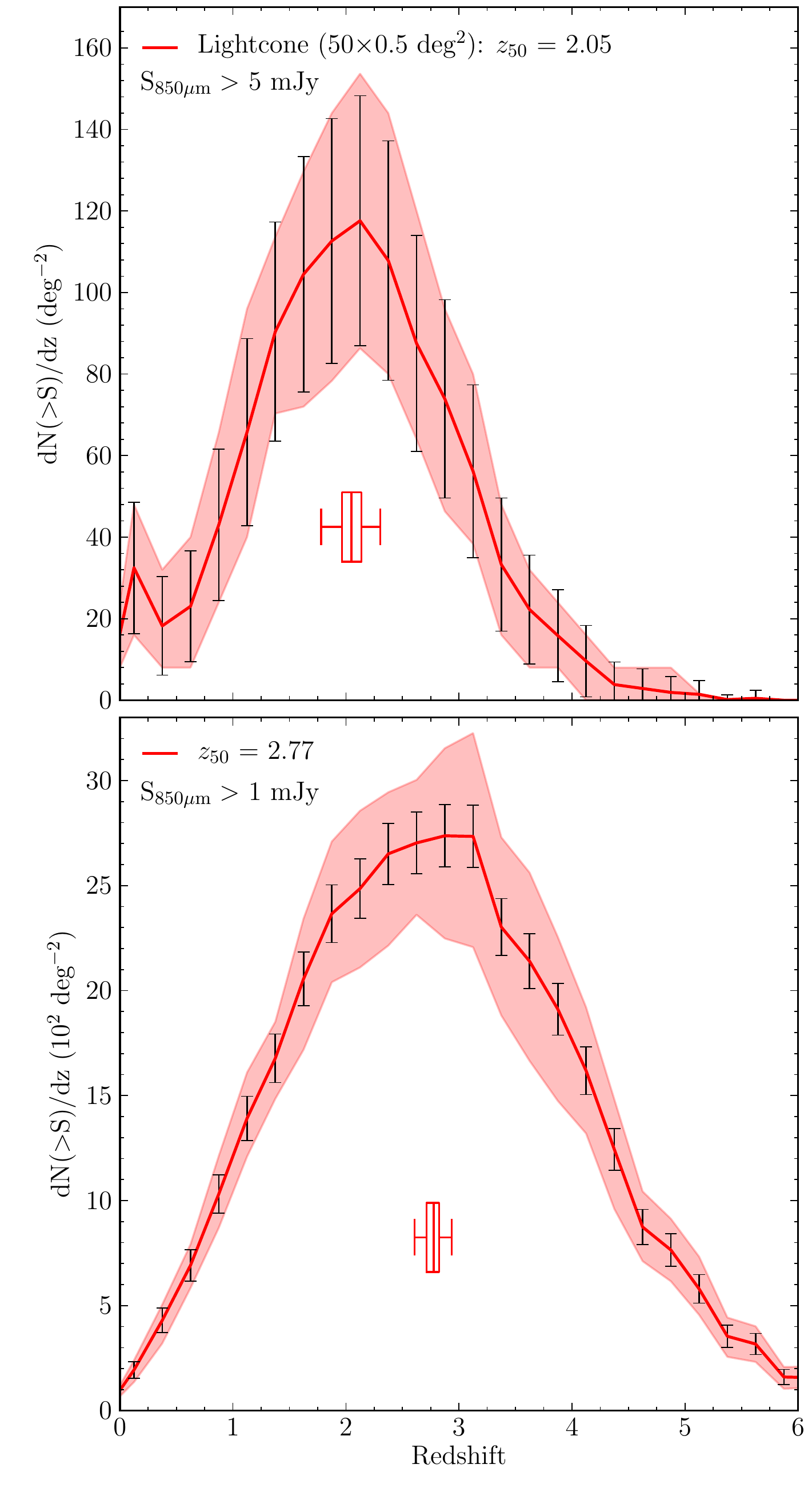}
\caption{The predicted redshift distribution for our $50\times0.5$ deg$^2$ fields for the flux limit indicated on each panel. The shaded red region shows the 16-84 ($1\sigma$) percentile of the distributions from the 50 individual fields. The solid red line is the distribution for the combined $25$ deg$^2$ field.  The boxplots represent the distribution of the median redshifts of the $50$ fields, the whiskers show the full range, with the box and central line indicating the inter-quartile range and median.  The errorbars show the expected $1\sigma$ variance due to Poisson errors.}
\label{fig:dndz_poisson}
\end{figure}

The redshift distributions for the `bright' $S_{850\mu\mathrm{m}}>5$ mJy and `faint' $S_{850\mu\mathrm{m}}>1$ mJy galaxy populations are shown in Fig. \ref{fig:dndz_poisson}.  
The shaded region shows the 16-84 (1$\sigma$) percentiles of the distributions from the 50 individual fields, arising from field-to-field variations. The errorbars indicate the $1\sigma$ Poisson errors.  The bright SMG population has a lower median redshift ($z_{50}=2.05$) than the faint one ($z_{50}=2.77$). We note that the median redshift appears to be a robust statistic with an inter-quartile range of 0.17 (0.11) for the bright (faint) population for the 0.5 deg$^2$ field size assumed.  The field-to-field variation seen in the bright population is comparable to the Poisson errors and thus random variations, whereas this field-to-field variation is greater compared to Poisson for the faint population.  In order to further quantify this field-to-field variance, we have performed the Kolmogorov-Smirnoff (K-S)
 test between the 1225 combinations of our 50 fields, for the bright and faint populations.  We find that for the bright population the distribution of $p$-values is similar to that obtained if we perform the same operation with 50 random samplings of the parent field, though with a slightly more significant low $p$-value tail. Approximately $10\%$ of the field pairs exhibit $p<0.05$, suggesting that it is not necessarily as uncommon as one would expect by chance to find that redshift distributions derived from non-contiguous pencil beams of sky fail the K-S test, as in \cite{Michalowksi12}.  For the faint population, $92\%$ of the field pairs have $p<0.05$.             

Thus, it appears that the bright population in the individual fields is more consistent with being a random sampling of the parent $25$ deg$^2$ distribution.  This is due to: (i) the number density of the faint population being $\sim 30$ times greater than the bright population, which significantly reduces the Poisson errors; and (ii) the median halo mass of the two populations remaining similar, $7.6$ $(5.5)$ $\times 10^{11}$ $h^{-1}$M$_{\odot}$ for our bright (faint) population implying that the two populations trace the underlying matter density with a similar bias.  We consequently predict that as surveys probe the SMG population down to fainter fluxes, we expect that they become more sensitive to field-to-field variations induced by large scale structure. 

\section{Comparison to ALESS}
\label{sec:ALESS}
In this section we make a detailed comparison of our model with observational data from the recent ALMA follow-up survey \citep{Hodge13} of LESS \citep{Weiss09}, referred to as ALESS.  LESS is an 870$\mu\mathrm{m}$ LABOCA (19.2$''$ FWHM) survey of 0.35 deg$^2$  (covering the full area of the ECDFS) with a typical noise level of $\sigma\sim1.2$ mJy/beam.  \cite{Weiss09} extracted 126 sources based on a S/N $>3.7\sigma$ ($\simeq S_{870\mu\mathrm{m}}>4.5$ mJy) at which they were $\sim70\%$ complete.  Of these 126 sources, 122 were targeted for cycle 0 observations with ALMA.  From these 122 maps, 88 were selected as `good' based on their rms noise and axial beam ratio, from which 99 sources were extracted down to $\sim1.5$ mJy.  The catalogue containing these 99 sources is presented in \cite{Hodge13}, with the resulting number counts and photometric redshift distribution being presented in \cite{Karim13} and \cite{Simpson13} respectively.  For the purposes of our comparison we randomly sample (without replacement) $70\%$ ($\sim 88/126$) of our $S_{850\mu\mathrm{m}}>4.5$ mJy sources from the central 0.35 deg$^2$ of our 50 mock maps\footnote{ We re-calculate the `effective' area of our follow-up surveys as $0.35$ deg$^2$ $\times N_{\rm{Good\,ALMA\,Maps}}/N_{\rm LESS\,Sources}\approx0.25$ deg$^2$ as in \cite{Karim13}}.  Around all of these sources we place $18''$ diameter masks ($\sim$ ALMA primary beam).  From these we extract  `follow-up' galaxies down to a minimum flux of  $S_{850\mu\rm m}=1.5$ mJy from the relevant lightcone catalogue.  We take into account the profile of the ALMA primary beam for this, modelling it as a Gaussian with an $18''$ FWHM, such that lightcone galaxies at a radius of 9$''$ from a source are required to be $>3$~mJy for them to be `detected.'   The result of this procedure is our `follow-up' catalogue.  We note that we do not attempted to simulate and extract sources from ALMA maps. 

\subsection{Number counts and source multiplicity}
\label{sec:ALESS_ncts}  
\begin{figure}
\centering
\includegraphics[width=\columnwidth]{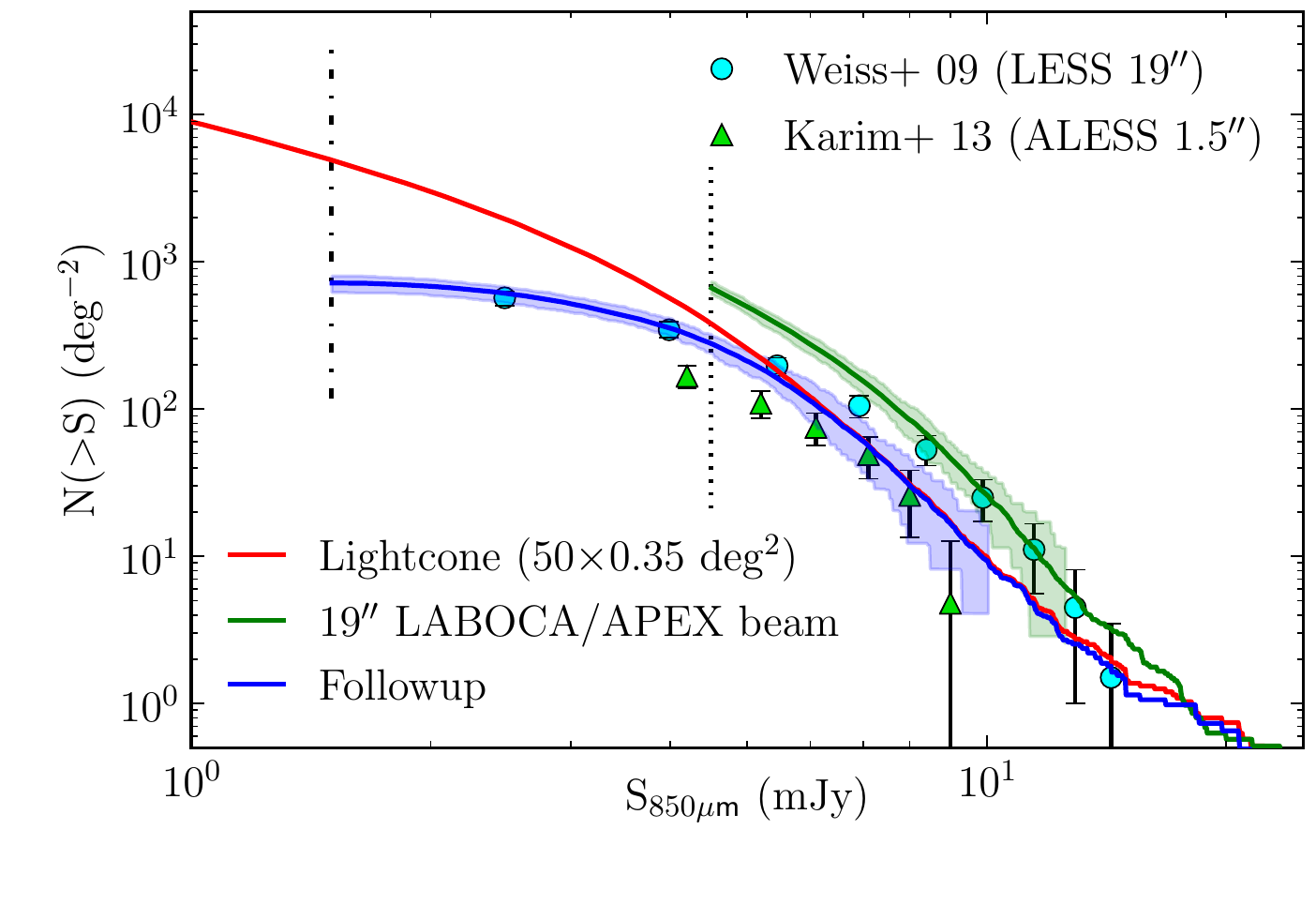}
\caption{Comparison with (A)LESS number counts.  The blue line is our prediction for our combined ($17.5$ deg$^2$) follow-up catalogues (described in text) and is to be compared to the ALESS number counts presented in Karim et al. (\citeyear{Karim13}; green triangles).  The green line is our $19''$ source-extracted number counts for the combined (17.5 deg$^2$) field and is to be compared to the number counts presented in Wei{\ss} (\citeyear{Weiss09}; cyan circles).  The shaded regions indicate the 10-90 percentiles of the distribution of the individual (0.35 deg$^2$) field number counts.  The red line is the number counts for the combined field from our lightcone catalogues.  The vertical dotted and dash-dotted lines indicate the 4.5 mJy single-dish source-extraction limit of LESS and the 1.5 mJy maximum sensitivity of ALMA respectively.}
\label{fig:ALESS_ncts}
\end{figure}

We present the number counts from our simulated follow-up catalogues in Fig. \ref{fig:ALESS_ncts} and observe a similar difference between our simulated single-dish and follow-up number counts as the (A)LESS survey found in their observed analogues (Wei{\ss} et al. 2009 and Karim et al. 2013 respectively).
Also evident is the bias inherent in our simulated follow-up compared to our lightcone catalogues at fluxes fainter than the source extraction limit of the single-dish survey. This arises because follow-up galaxies are only selected due to their on-sky proximity to a single-dish source, so they are not representative of a blank-field population.  For this reason \cite{Karim13} do not present number counts fainter than the source extraction limit of LESS, despite the ability of ALMA to probe fainter fluxes.  Whilst our model agrees well with both interferometric and single-dish data at bright fluxes,  as discussed in Section \ref{sec:nctsbeam}, our single-dish predictions are in excess of the \cite{Weiss09} data at fainter fluxes ($S_{850\mu\rm m}\lesssim 7$~mJy). We also observe a minor excess in our `follow-up' number counts when compare to the \cite{Karim13} data for $S_{850\mu\rm m}\lesssim5$~mJy.  

\begin{figure}
\centering
\includegraphics[width=\columnwidth]{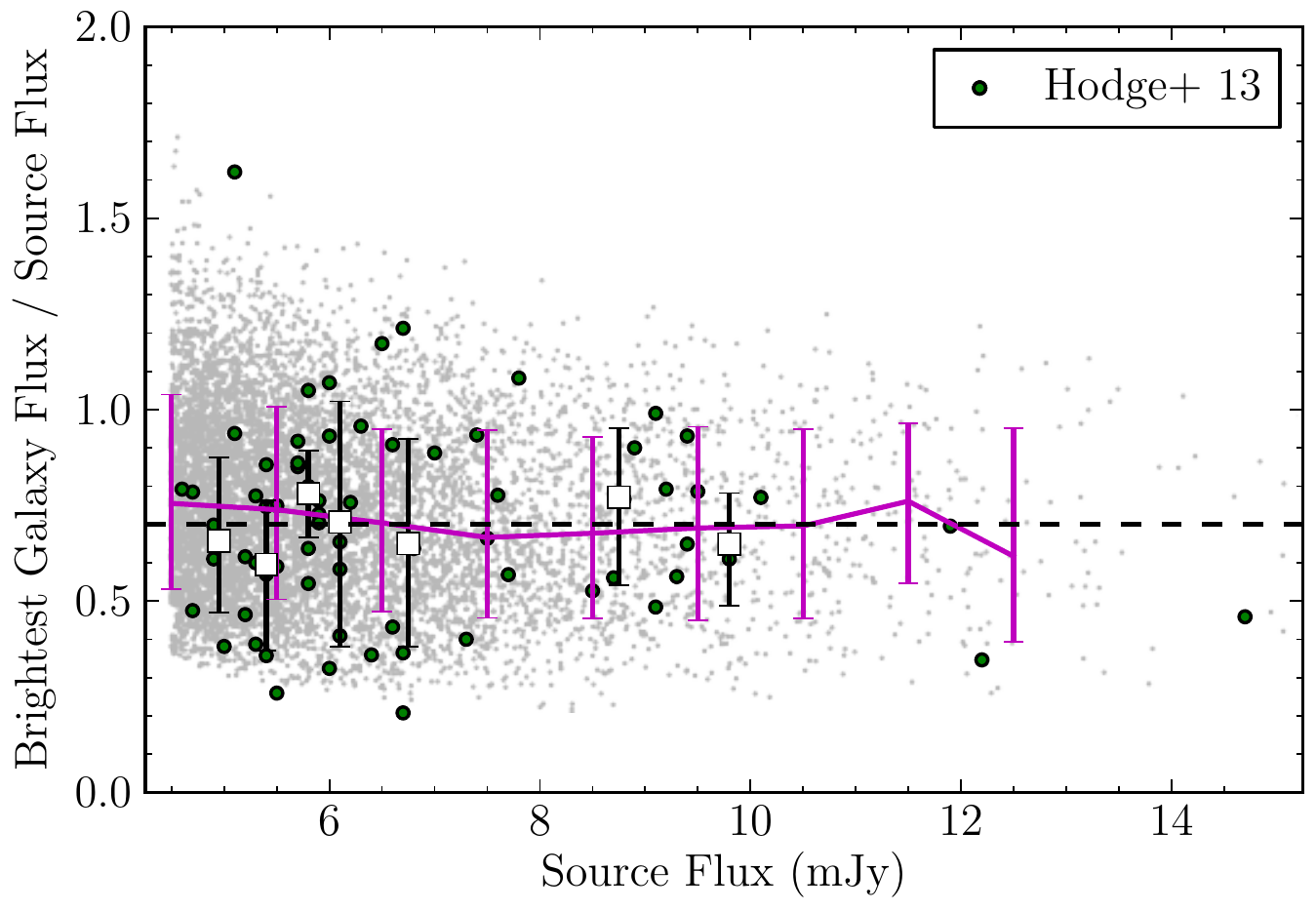}
\caption{Ratio of brightest galaxy component flux to single-dish source flux.  Grey scatter points show the brightest galaxies from our targeted sources over the combined 17.5 deg$^2$ simulated field.  The magenta line shows the median in a given flux bin. Observational data is taken from the Hodge et al. (\citeyear{Hodge13}) ALESS catalogue.  The white squares indicate the median observational flux ratio and source flux in a given bin, with the binning chosen such that there are roughly equal numbers of sources in each bin.  Error bars indicate the $1\sigma$ percentiles of the ratio distribution in a given flux bin for both simulated and observed data.  The black dashed line is a reference line drawn at $70\%$.}
\label{fig:component_flux}
\end{figure}

\begin{figure}
\centering
\includegraphics[width=\columnwidth]{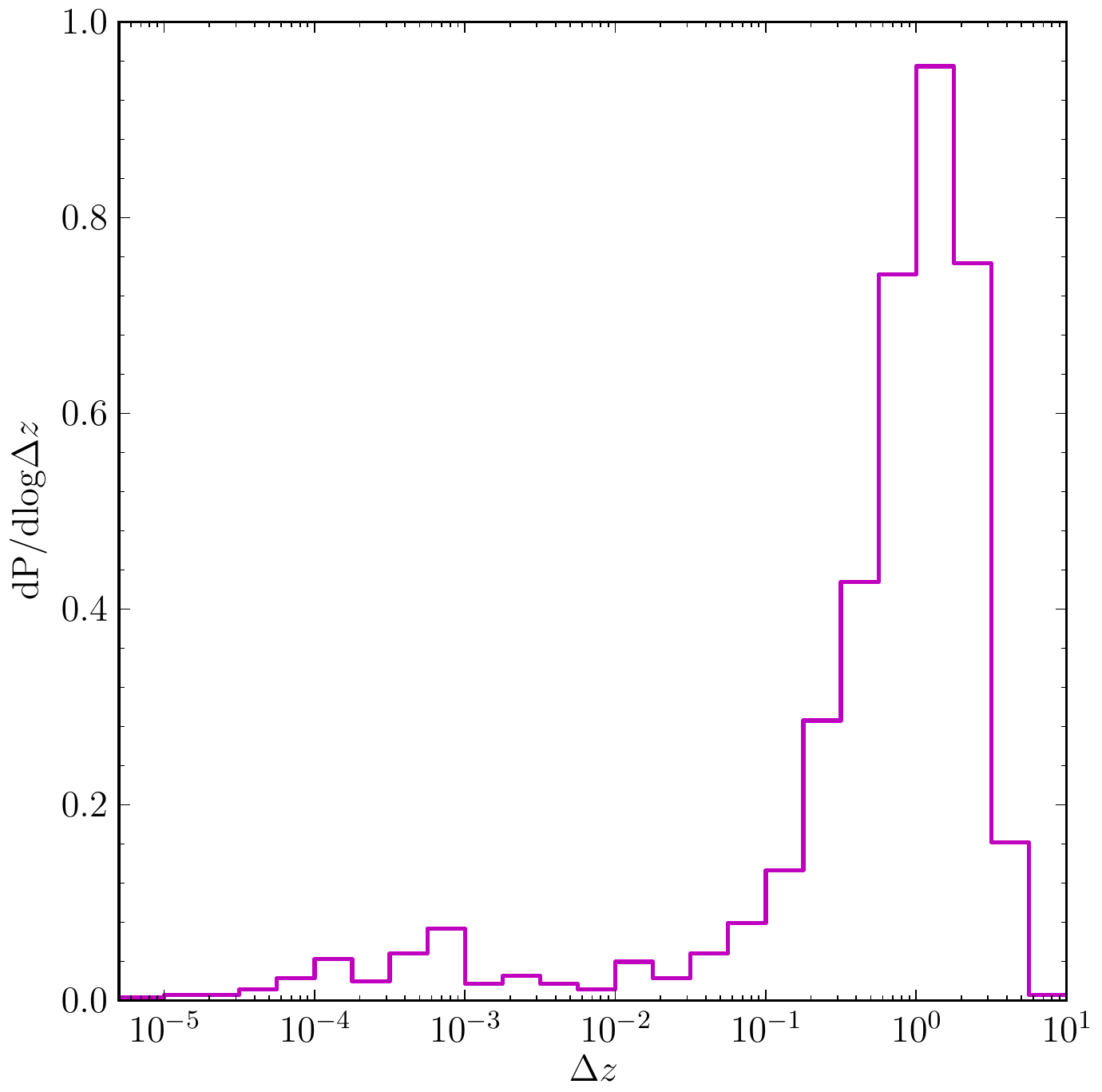}
\caption{Distribution of the logarithm of redshift separation of the brightest two ALMA components of a $S_{850 \mu \rm m}>4.5$ mJy single-dish source for our combined (17.5 deg$^2$) field.}
\label{fig:ALESS_delta_z}
\end{figure}

We show the ratio of the brightest follow-up galaxy flux for each source to the source flux in Fig. \ref{fig:component_flux} and our prediction is in excellent agreement with the observed sample, with the brightest of our follow-up galaxies being roughly $70\%$ of the source flux on average.
This fraction is approximately constant over the range of source fluxes probed by LESS.  The scatter of our simulated data is also comparable to that seen observationally.  Not plotted in Fig. \ref{fig:component_flux} are sources for which the brightest galaxy is below the flux limit of ALMA. These account for $\sim10\%$ of our sources.  \cite{Hodge13} found that $\sim21\pm5\%$ of the $88$ ALMA `Good Maps' yielded no ALMA counterpart.  The greater fraction of blank maps in the observational study could be caused by extended/diffuse SMGs falling below the detection threshold of ALMA and/or a greater source multiplicity in the observed sample.  We present a breakdown of the predicted ALMA multiplicity of our simulated LESS sources compared to the observed \cite{Hodge13} sample in Table \ref{tab:ALMAtable}.  Our simulated follow-up catalogue is consistent with the observed sample at $\sim2\sigma$. However, we caution that it is difficult to draw strong conclusions from this comparison due to the relatively small number of observed sources.  We also note that we observe a similar trend for increasing source multiplicity with flux to that suggested in \cite{Hodge13}.  For example, at $S_{850\mu\rm m}=5$~mJy the fraction of simulated sources with 2 ALMA components is $\sim10\%$ increasing to $\sim40\%$ at $S_{850\mu\rm m}=10$~mJy with the fraction of simulated sources with 1 ALMA component decreasing from $\sim70\%$ to $\sim60\%$ over the same flux range.  This is in contrast to conclusions drawn from Fig. \ref{fig:multiplicity} and shows that this observed trend is probably caused by the flux limit of the interferometer, meaning that faint components are undetected. 
\begin{table}
\centering
\caption{A breakdown of the number of ALMA components from our simulated sample for comparison with the observed sample of Hodge et al. (\citeyear{Hodge13}).  The columns are: (1) the number of ALMA components; (2) the percentage of our simulated sources with that number of ALMA components; (3) the percentage of observed LESS sources with `good' ALMA maps that contain that number of ALESS components, errors are Poisson; and (4) the number of observed LESS sources with `good' ALMA maps that contain that number of ALMA components.}
\label{tab:ALMAtable}
\begin{tabular}{cccc}\hline
$N$            & Sim. (\%)& Obs. (\%)& Obs. (/88) \\ \hline
0                    & 10.6         & $22\pm5$      &      19               \\
1                    & 72.2         & $51\pm8$      &		45               \\
2                    & 16.5         & $22\pm5$      &		19               \\
3                    & 0.70         & $5\pm3$       &		 4               \\
4                    & 0.01         & $1\pm1$       &		 1               \\ \hline
\end{tabular}

\end{table}  
     
For comparison with future observations we calculate $\Delta z$ for all of our sources with $\geq2$ ALMA components as the redshift separation of the brightest two.  We show the resulting distribution in Fig. 12.  It is of a similar bimodal shape to the distribution presented in Fig. 8 and supports the idea that, in our model, blended galaxies are predominantly chance line of sight projections with a minor peak at small $\Delta z$ due to clustering.  We leave this as a prediction for future spectroscopic redshift surveys of interferometer identified SMGs (e.g. Danielson et al. in prep).

\subsection{Redshift distribution}
\begin{figure}
\centering
\includegraphics[width=\columnwidth]{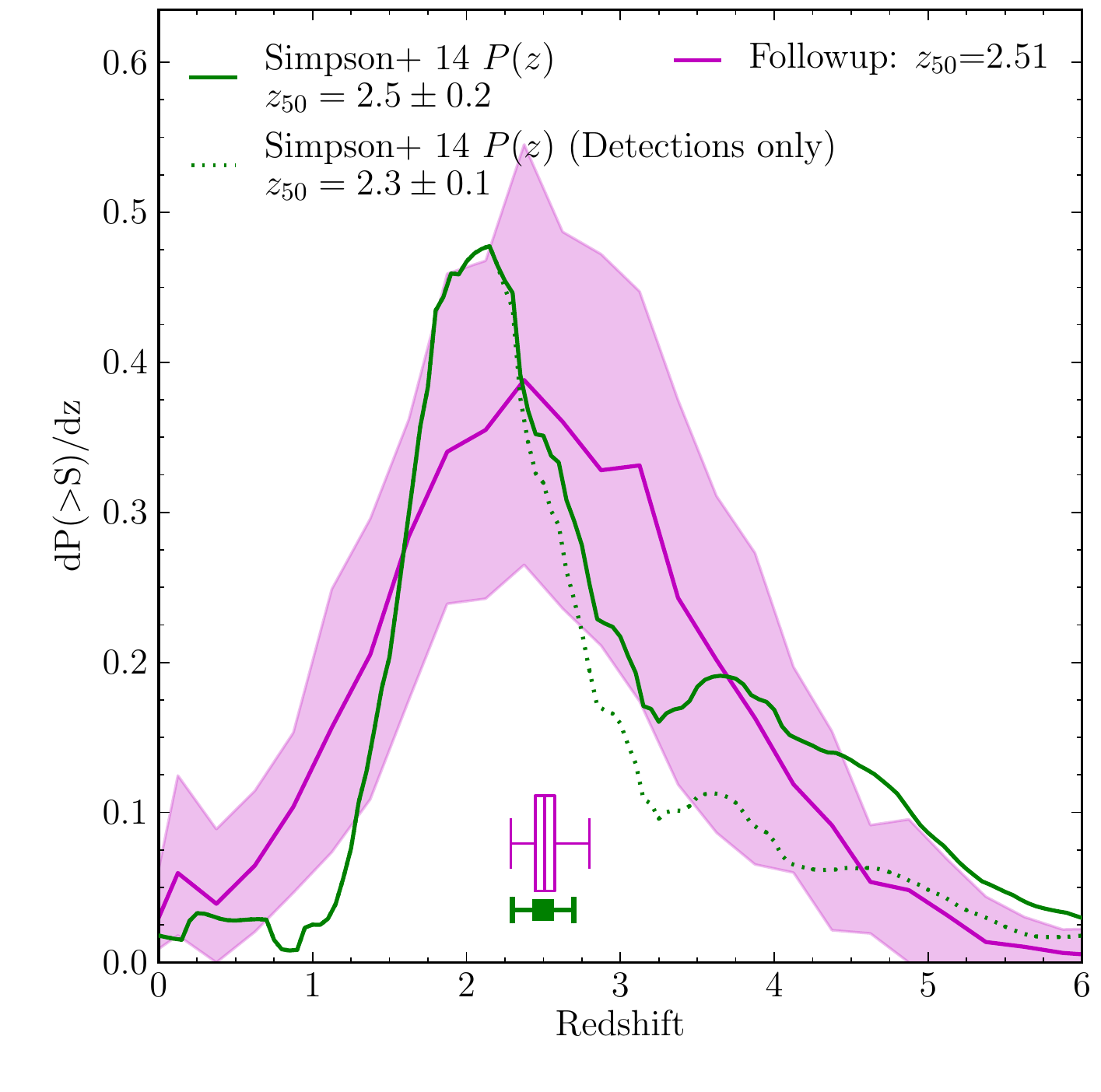}
\caption{Comparison of normalised redshift distributions for the simulated and observed ALESS surveys.  We show the Simpson et al. (\citeyear{Simpson13}) $P(z)$, the sum of the photometric redshift probability distributions of each galaxy, both including redshifts derived from $H$-band absolute magnitude modelling for `non-detections' (see Simpson et al. for details, solid green line) and for photometric detections only (dotted green line).  The square marker indicates the observed median redshift (including $H$-band modelled redshifts), with associated errors.  The magenta solid line is the distribution for the simulated, combined 17.5 deg$^2$ field with the shaded region showing the 10-90 percentiles of the distributions from the 50 individual fields.  The boxplot shows the distribution of median redshifts for each of the 50 individual fields, the whiskers indicate the full range, with the box and line indicating the inter-quartile range and median respectively.} 
\label{fig:ALESS_dpdz}
\end{figure}

One of the main advantages of the 99 ALMA sources identified in \cite{Hodge13} is that the greater positional accuracy ($\sim0.2''$) provided by ALMA allows accurate positions to be determined without introducing biases associated with selection at wavelengths other than sub-mm (e.g. radio).  \cite{Simpson13} derived photometric redshifts for 77 of 96 ALMA SMGs\footnote{Three of the 99 SMGs presented in \cite{Hodge13} lay on the edge of ECDFS with coverage in only two IRAC bands, and so were not considered further in \cite{Simpson13}.}.  The remaining 19 were only detected in $\leq3$ bands and so reliable photometric redshifts could not be determined.  Redshifts for these `non-detections' were modelled in a statistical way based on assumptions regarding the $H$-band absolute magnitude ($M_{H}$) distribution of the 77 `detections' \citep[see][for more details]{Simpson13}.  We compare the redshift distribution presented in \cite{Simpson13}   to that of our simulated follow-up survey in Fig. \ref{fig:ALESS_dpdz}.  For the purposes of this comparison we have included the $P(z)$, the sum of the photometric redshift probability distributions for each galaxy, with (solid green line) and without (dotted green line) the $H$-band modelled redshifts.             

Our model exhibits a high redshift ($z>4$) tail when compared to the top panel of Fig. \ref{fig:dndz_poisson}, due to the inclusion of fainter galaxies in this sample, and is in excellent agreement with the median redshift of the observed distribution.  We performed the K-S test between each of our 50 follow-up redshift distributions and the ALESS distribution and find a low median $p$ value of $0.16$ with $18\%$ of the K-S tests exhibiting $p<0.05$.  We do note, however, that the $M_{H}$ band modelling of the 19 `non-detections' ($\sim 20\%$ of the sample), and the sometimes significant photometric errors may affect the  observed distribution.  

\begin{figure}
\centering
\includegraphics[width=\columnwidth]{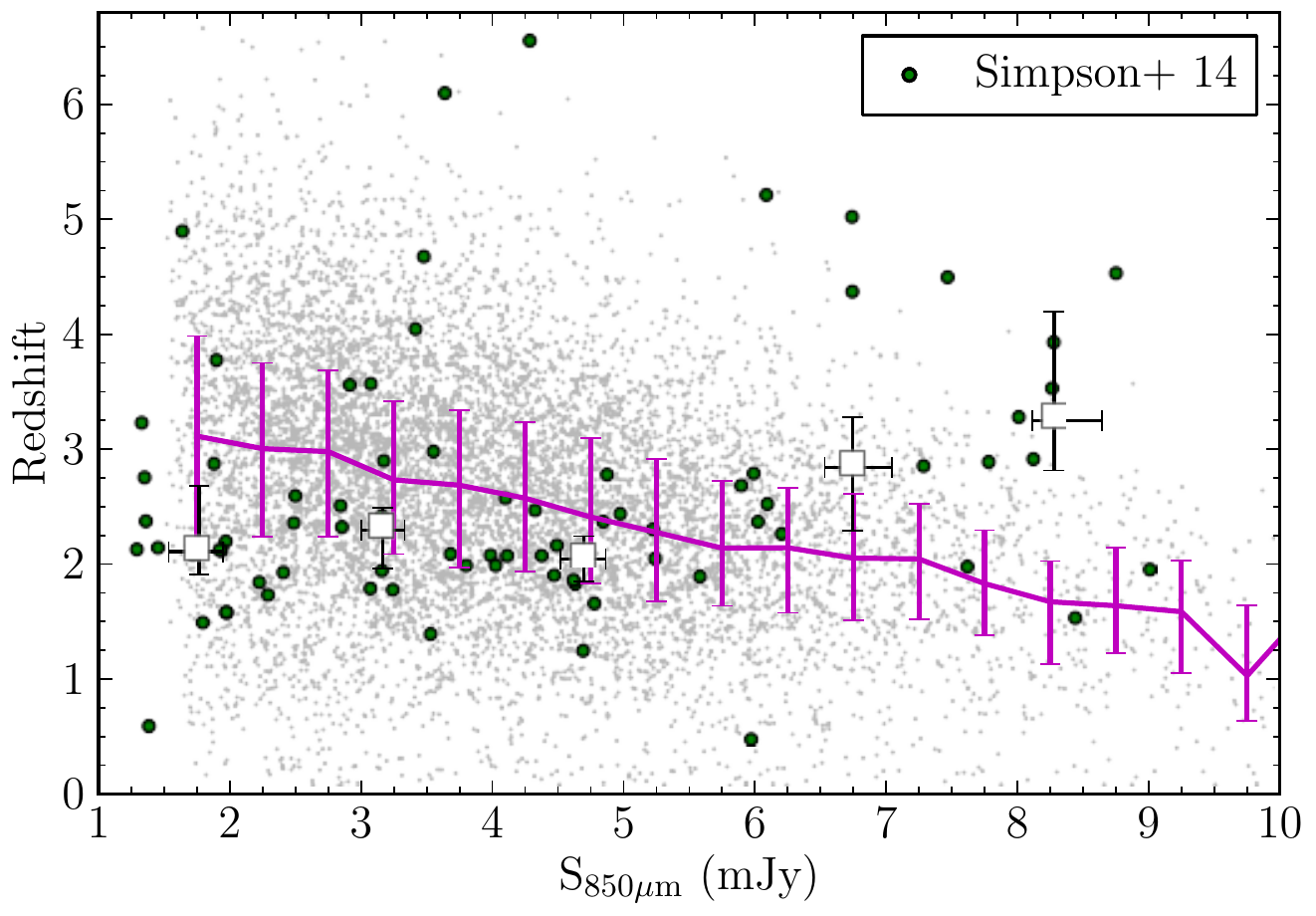}
\caption{Relation between $S_{850\mu\mathrm{m}}$ and redshift for our simulated follow-up galaxies over our combined 17.5 deg$^2$ field.  Solid line shows the median redshift in a given $1$ mJy $S_{850\mu\mathrm{m}}$ bin with errorbars indicating the inter-quartile range.  Observational data from Simpson et al. (\citeyear{Simpson13}) has been binned in $2$ mJy bins, with the median redshift plotted as the white squares with errorbars indicating $1\sigma$ bootstrap errors.} 
\label{fig:z_vs_S}
\end{figure}

We also investigate whether or not our model reproduces the same behaviour as seen in ALESS between redshift and $S_{850\mu\mathrm{m}}$ in Fig. \ref{fig:z_vs_S}.  Our model predicts that at lower redshift our simulated SMG population is generally brighter whilst in the observational data the opposite appears to be the case. However, \cite{Simpson13} argue that this trend in their data is not significant and that their non-detections, 14/19 of which are at $S_{870\mu\mathrm{m}}<2$ mJy, would most likely render it flat if redshifts could be determined for these galaxies.  
    
\section{Multi-wavelength surveys}
\label{sec:multi_lam}
Until now we have focussed on surveys performed at 850 $\mu$m, traditionally the wavelength at which most sub-mm surveys have been performed.  However, there are now a number of observational blank-field surveys performed at other sub-mm wavelengths \citep[e.g.][]{Scott12,Chen13,Geach13}.  In this section we briefly investigate the effects of the finite single-dish beam-size at 450 $\mu$m ($\sim 8''$ FWHM e.g. SCUBA2/JCMT) and 1100 $\mu$m ($\sim 28''$ FWHM e.g. AzTEC/ASTE\footnote{A\emph{z}tronomical Thermal Emission Camera/Atamaca Sub-millimetre Telescope Experiment}).  We add that due to our self-consistent dust model the results presented in this section are genuine multi-wavelength predictions and do not rely on applying an assumed fixed flux ratio\footnote{At 450 $\mu$m galaxies at high redshift ($z\gtrsim5.5$) have $\lambda_{\rm rest}<70$ $\mu$m and therefore the sub-mm flux calculated by our dust model may be systematically incorrect when compared to \grasil predictions (see Section \ref{sec:dust_model}).  We expect the contribution of such galaxies to our $450$ $\mu$m population to be small.}.

We create lightcones as described in Section \ref{sec:create_mock_surveys}, taking the lower flux limit at which we include galaxies in our lightcone catalogue as the limit above which $90\%$ of the EBL is resolved at that wavelength, as predicted by our model.  This is $0.125$ ($0.02$) mJy at $450$ ($1100$) $\mu$m.  As at 850 $\mu$m, our EBL predictions are in excellent agreement with observational data from the \emph{COBE} satellite.  At $450$ ($1100$) $\mu$m we predict a background of $140.1$ ($23.9$) Jy deg$^{-2}$ compared to $142.6^{+177.1}_{-102.4}$ ($24.8^{+26.5}_{-20.8}$) Jy deg$^{-2}$ found observationally by Fixsen et al. (1998).  We  follow the same procedure as described in Section \ref{sec:create_submm_maps} for creating our mock maps.  However, we change the standard deviation of our Gaussian white noise such that the match-filtered noise-only maps have a $\sigma$ of $\sim 4$ $(1)$ mJy/beam at 450 (1100) $\mu$m to be comparable to published blank-field surveys at that wavelength \citep[e.g.][]{Aretxaga11,Casey13}.  

\begin{figure*}
\centering
\includegraphics[width=\linewidth]{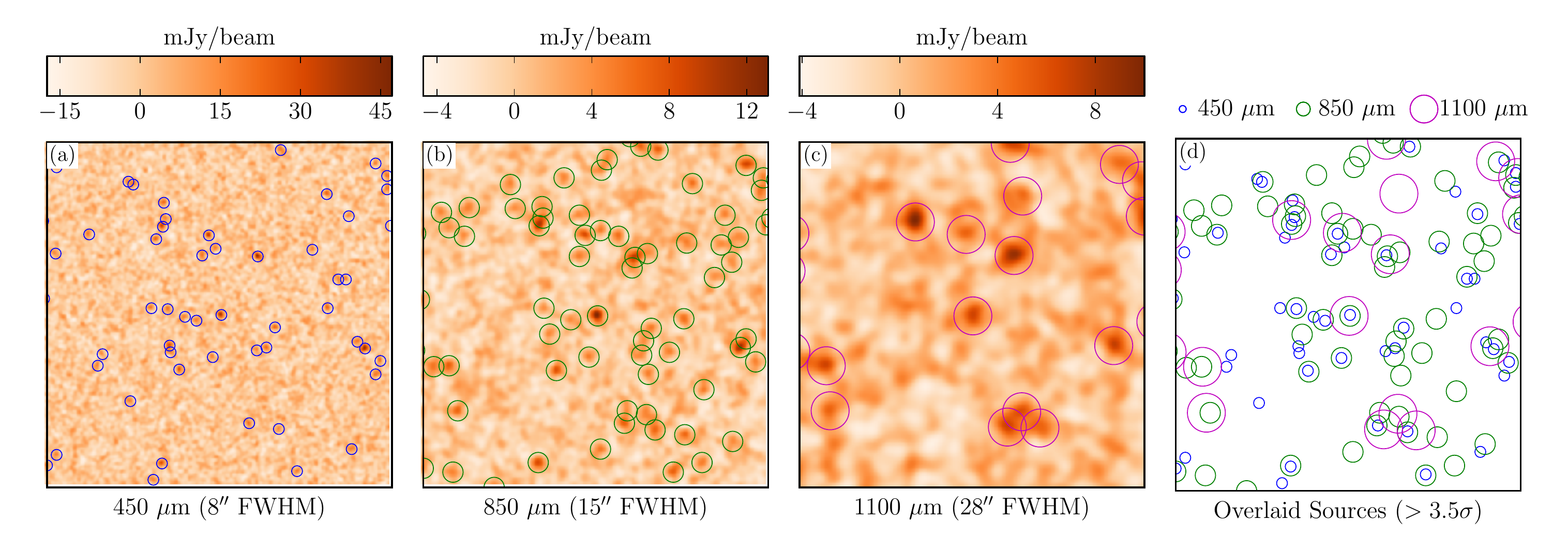}
\caption{Thumbnails of the same $0.2\times0.2$ deg$^2$ area as depicted in panels (a)-(d) of Fig. \ref{fig:thumbs} but at (a) $450~\mu$m, (b) $850~\mu$m and (c) $1100~\mu$m.  Overlaid are the $>3.5\sigma$ sources, as circles centred on the source position with a radius of $\sqrt{2}\times$FWHM of the telescope beam at that wavelength.  In (d) the $>3.5\sigma$ sources at each wavelength are overlaid, without background for clarity.}
\label{fig:multi_lam_thumb}
\end{figure*}

Thumbnails of the same area, but for different wavelength maps, are shown for comparison in Fig. \ref{fig:multi_lam_thumb}. 
The effect of the beam size increasing with wavelength is clearly evident, as is the resulting multiplicity of some of the sources.  Drawing physical conclusions from this source multiplicity is  not trivial. Selection at shorter wavelengths tends to select lower redshift and/or hotter dust temperature galaxies.  For example, for an arbitrary flux limit of 1 mJy the median redshifts of the $450$, $850$ and $1100~\mu$m populations in our model are $2.31$, $2.77$ and $2.93$ respectively. This is complicated further by the fact that, as we have shown in this paper, at sub-mm wavelengths single-dish detected sources are likely to be composed of multiple individual galaxies, which may (or may not) also be bright at other wavelengths depending on the SED of the object, and that these galaxies are generally physically unassociated.  If we restrict our analysis to galaxies only, thus avoiding complications caused by the single-dish beam, and consider flux limits of 12, 4 and 2 mJy at 450, 850 and 1100 $\mu$m respectively\footnote{These flux limits were motivated by the median flux ratios of our lightcone galaxies of $S_{1100\mu \rm m}/S_{850\mu \rm m}\approx 0.5$ and $S_{850\mu \rm m}/S_{450\mu \rm m}\approx 0.3$} we find median redshifts of $1.71$, $2.26$ and $2.55$ for selection at each wavelength respectively.  If we now consider a sample that satisfy these selection criteria at all wavelengths we find a median redshift of $z=2.09$, and that this sample comprises $52$, $80$ and $66\%$ of the single band selected samples at $450$, $850$ and $1100~\mu$m respectively.  It is unsurprising that the multi-wavelength selected sample overlaps most with the intermediate $850~\mu$m band. 

\begin{figure}
\centering
\includegraphics[width=\columnwidth]{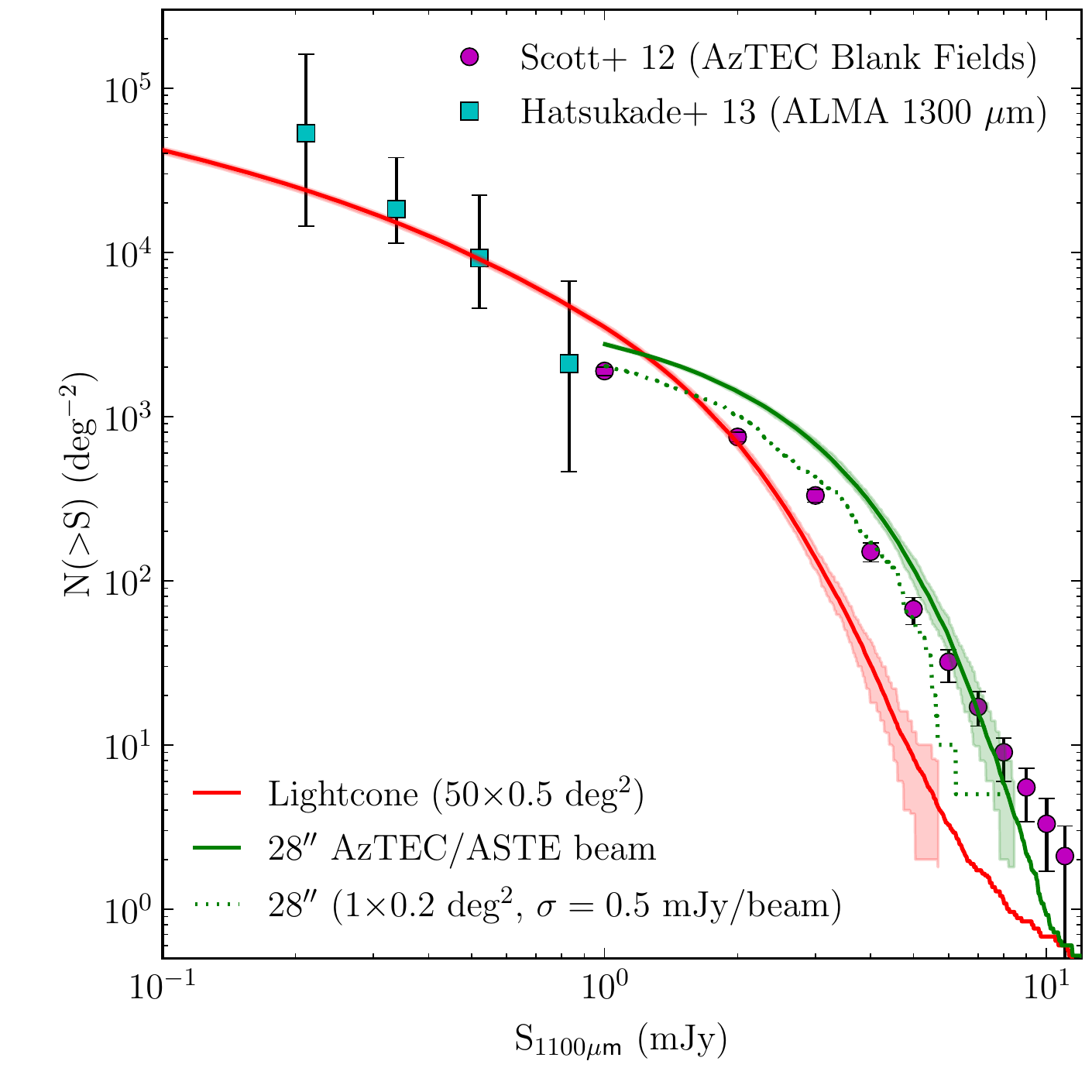}
\caption{Predictions for cumulative blank-field single-dish number counts at 1100 $\mu$m.  Number counts from our lightcone (red line) and $28''$ FWHM beam ($\sigma = 1$~mJy/beam) source-extracted (green solid line) catalogues are shown.  The shaded regions are the 10-90 percentiles of our individual field number counts.  We also show number counts derived from a smaller field with Gaussian white noise of $\sigma = 0.5$~mJy/beam (green dotted line).  Blank field single-dish observational data is taken from Scott et al. (\citeyear{Scott12}; magenta circles) and serendipitous ALMA 1300~$\mu$m number counts from Hatsukade et al. (\citeyear{Hatsukade13}; cyan squares) assuming $S_{1300\mu\rm m}/S_{1100\mu\rm m} = 0.71$.}
\label{fig:AzTEC_ncts}
\end{figure}

In Fig. \ref{fig:AzTEC_ncts} we present the $1100~\mu$m number counts from our source-extracted and lightcone catalogues.
The observational data from \cite{Scott12} is a combined sample of previously published blank field single-dish number counts from surveys of varying area and sensitivity with a total area of $1.6$ deg$^2$, $1.22$ deg$^2$ of which were taken using using the AzTEC/ASTE configuration.  As at $850$ $\mu$m, considering the effects of the finite beam-size brings the model into better agreement with the single-dish observational data. We also plot, from \cite{Hatsukade13},  $1300~\mu$m number counts derived from serendipitous detections found in targeted ALMA observations of  star-forming galaxies at $z\sim1.4$ (converted to $1100~\mu$m counts assuming $S_{1300\mu\rm m}/S_{1100\mu\rm m} = 0.71$ as is done in Hatsukade et al.).  These benefit from the improved angular resolution of the ALMA instrument $\sim 0.6-1.3''$ FWHM and can thus probe to fainter fluxes than the single-dish data.  Due to the higher angular resolution of these observations they are to be compared to the lightcone catalogue number counts (red line) and show good agreement with our model.  However, we caution that due to the targeted nature of the Hatsukade et al. observations they may not be an unbiased measure of a blank field population.  As the Scott et al. (2012) counts are derived from multiple fields of varying area and sensitivity, we also show in Fig. \ref{fig:AzTEC_ncts} number counts derived from a single 0.2 deg$^2$ field which has matched-filtered noise of $0.5$~mJy/beam (green dotted line), similar to the 1100~$\mu$m counts from the SHADES fields \citep{Hatsukade11} used in the \cite{Scott12} sample.  This shows better agreement with the Scott et al. data in the range $1\lesssim S_{1100\mu\rm m}\lesssim5$ mJy (at brighter fluxes the smaller field will suffer from a lack of bright objects) which leads us to the conclusion that the discrepancy between our $\sigma = 1$~mJy/beam number counts (green solid line) and the \cite{Scott12} data is due more to our assumed noise than of physical origin.  As instrumental/atmospheric noise is unlikely to be Gaussian white noise in real observations, and various methods are used in filtering the observed maps to account for this, which we do not model here, we consider further investigation of the effect of such noise on observations beyond the scope of this work.  At $\gtrsim5$~mJy our $\sigma = 1$~mJy/beam, 0.5~deg$^2$ number counts (solid green line) agree well with the \cite{Scott12} data, as the field size is more comparable to the largest field used in Scott et al. (0.7~deg$^2$), and instrumental noise will have less of an effect on both the simulated and observational data.

\begin{figure}
\centering
\includegraphics[width=\columnwidth]{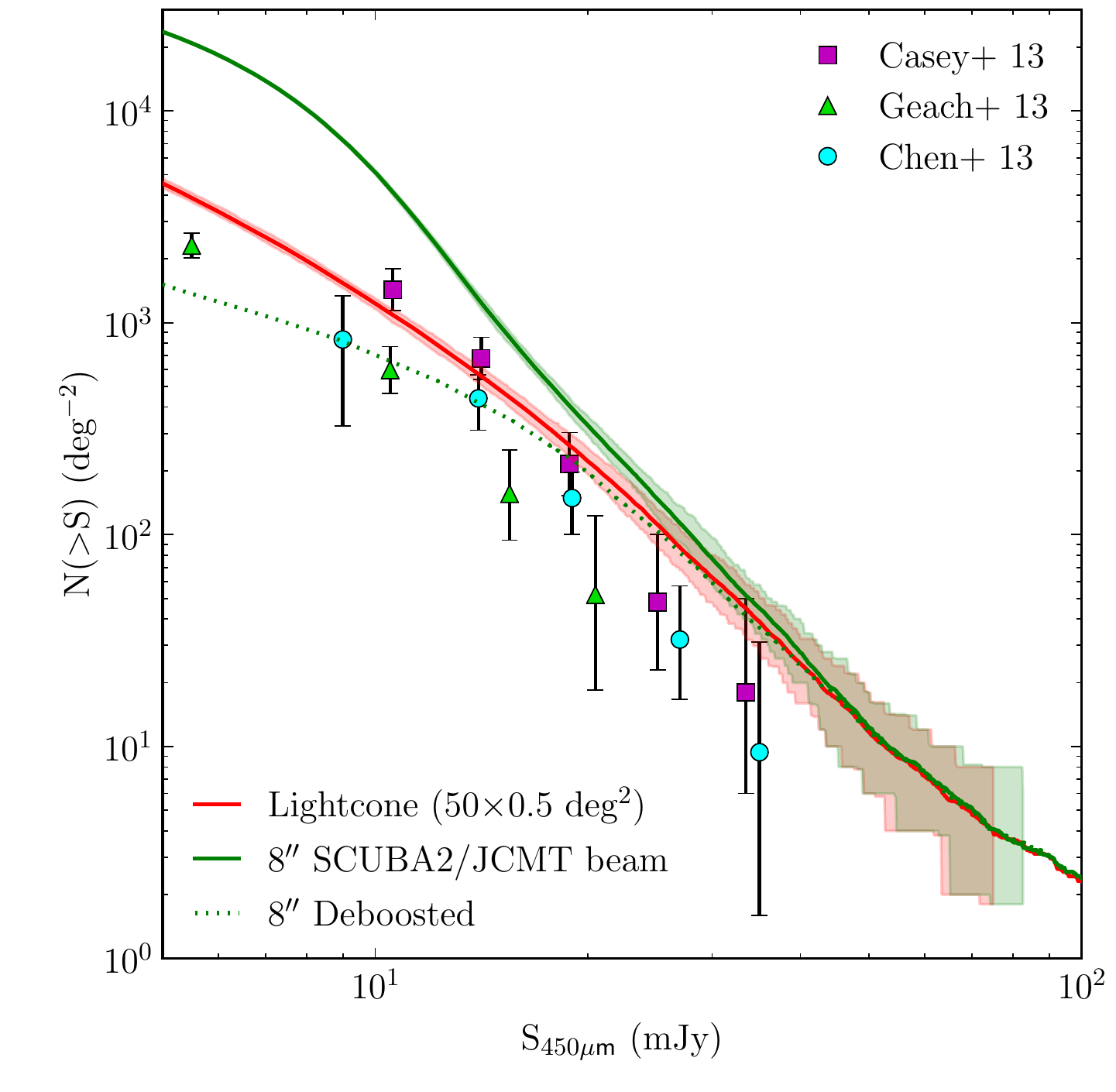}
\caption{Predictions for cumulative blank-field single-dish number counts at 450~$\mu$m.  Number counts from our lightcone (red) and $8''$ FWHM beam ($\sigma = 4$~mJy/beam) source-extracted (green) catalogues are shown for our combined $25$~deg$^2$ field.  The dotted green line shows the de-boosted source-extracted counts for the combined field (see text).  The shaded regions show the 10-90 percentiles of our individual field number counts.  Observational data is taken from Casey et al. (\citeyear{Casey13}; magenta squares), Geach et al. (\citeyear{Geach13}; green triangles) and Chen et al. (\citeyear{Chen13}; cyan circles).}
\label{fig:450ncts}
\end{figure}

\begin{figure}
\centering
\includegraphics[width=\columnwidth]{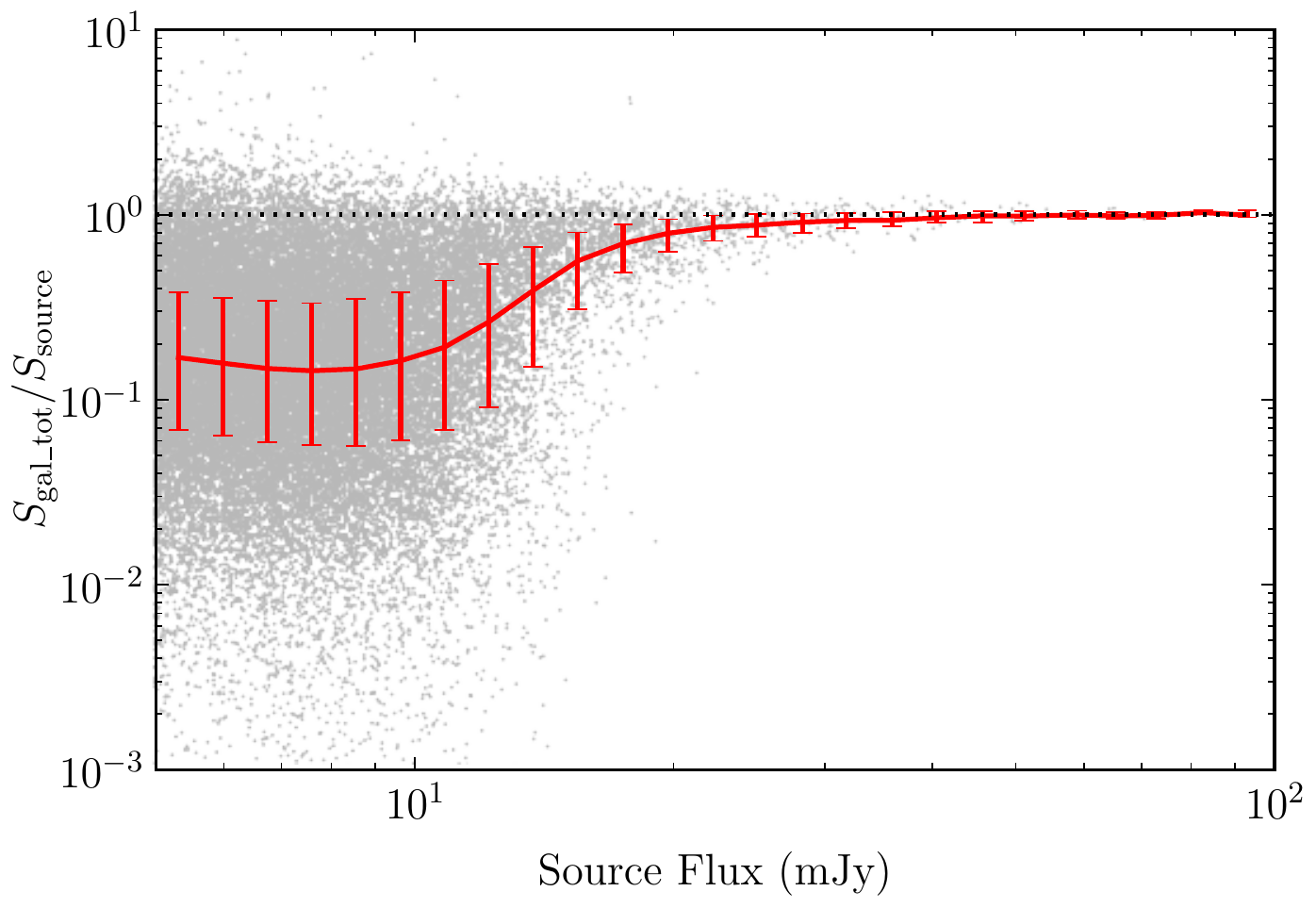}
\caption{Ratio of total galaxy flux (see Section \ref{sec:multiplicity}) to source flux at $450$~$\mu$m.  Red line and errorbars shows median and inter-quartile range in a given logarithmic flux bin respectively. For clarity, only 5\% of sources have been plotted as grey dots.}
\label{fig:450db}
\end{figure}

The number counts at $450~\mu$m are presented in Fig. \ref{fig:450ncts}.
We attribute the enhancement in our simulated source-extracted counts at $S_{450\mu\rm m}\sim8$~mJy to Eddington bias caused by the instrumental noise rather than an effect of the $8''$ beam.  In order to account for this we `deboost' our $S_{450\mu\rm m}>5$ mJy sources following a method similar to one outlined in \cite{Casey13}.  The total galaxy flux of each of our $S_{450\mu \rm m}>5$ mJy sources is calculated as described in Section \ref{sec:multiplicity} and we plot this as a ratio of source flux in Fig. \ref{fig:450db}.  
We multiply the flux of our 450 $\mu$m sources by the median of this ratio (red line) before re-calculating the number counts (green dotted line in Fig. \ref{fig:450ncts}).  These corrected number counts show good agreement with observational data in the flux range $5\lesssim S_{450\mu\rm m}\lesssim20$ but may slightly overestimate the counts for $S_{450\mu\rm m}\gtrsim20$.    

\section{Summary}
\label{sec:Summary}
We present predictions for the effect of the coarse angular-resolution of single-dish telescopes, and field-to-field variations, on observational surveys of SMGs.  An updated version of the \galform semi-analytic galaxy formation model is coupled with a self-consistent calculation for the reprocessing of stellar radiation by dust in order to predict the sub-mm emission from the simulated galaxies.  We use a sophisticated lightcone method to generate mock catalogues of SMGs out to $z=8.5$, from which we create mock sub-mm maps replicating observational techniques.  Sources are extracted from these mock maps to generate our source-extracted catalogue and show the effects of the single-dish beam on the predicted number counts.  To ensure a realistic background in our maps, we include model SMGs down to the limit above which $90\%$ of our total predicted EBL is resolved.  Our model shows excellent agreement with EBL observations from the \emph{COBE} satellite at 450, 850 and 1100~$\mu$m.  We generate $50\times 0.5$ deg$^2$ randomly orientated surveys to investigate the effects of field-to-field variations.  

The number counts from our $850$ $\mu$m source-extracted catalogues display a significant enhancement over those from our lightcone catalogues at brighter fluxes ($S_{850\mu\rm m}>1$ mJy) due to the sub-mm emission from multiple SMGs being blended by the finite single-dish beam into a single source.  The field-to-field variations predicted from both lightcone and source-extracted catalogues for the $850$ $\mu$m number counts are comparable to or less than quoted observational errors, for simulated surveys of $0.5$ deg$^2$ area with a $15''$ FWHM beam ($\sim$ SCUBA2/JCMT). Quantitatively we predict a field-to-field variation of 0.34 dex at $S_{850 \mu\rm m}=10$ mJy in our source-catalogue number counts.  Typically $\sim3{-}6$ galaxies to contribute $90\%$ of the galaxy flux of an $S_{850 \mu \rm m}=5$ mJy source, and this multiplicity slowly decreases with increasing flux over the range of fluxes investigated by blank-field single-dish surveys at $850$ $\mu\mathrm{m}$.  We find further that these blended galaxies are mostly physically unassociated, i.e. their redshift separation implies that they are chance projections along the line of sight of the survey.

Our redshift distributions predict a median redshift of $z_{50}=2.0$ for our `bright' ($S_{850\mu\mathrm{m}}>5$ mJy) galaxy population and $z_{50}=2.8$ for our `faint' ($S_{850\mu\mathrm{m}}>1$ mJy) galaxy population.  We leave these as predictions for blank field interferometric surveys of comparable area.  We also observe that the field-to-field variations we predict for our bright population are comparable to those expected for Poisson errors, whereas for our faint population the field-to-field variations are greater than Poisson.

A comparison between the ALESS survey and our model reveals that the model can reproduce the observed difference between observed single-dish and interferometer number counts, as well as estimates for the multiplicity of single-dish sources consistent (at $\sim2\sigma$) with those derived observationally.  It is in excellent agreement with observed relations between the flux of the brightest interferometric counterpart of a source and the source flux. The model also reproduces the median redshift of the observed photometric redshift distribution.  In addition, we predict that the majority of the interferometric counterparts are physically unassociated, and leave this as a prediction for future spectroscopic redshift surveys of such objects.

We also present predictions for our lightcone and source-extracted catalogue number counts at $450$ and $1100~\mu$m, which show good agreement with the observational data .  It is evident that the finite beam-size does not lead to a significant enhancement of the number counts at $450$, as opposed to $850$ and $1100~\mu$m, as the beam-size at $450~\mu$m is significantly smaller.  At $1100~\mu$m we show that the model agrees well with both interferometric and single-dish observational number counts.  Due to our dust model these are genuine multi-wavelength predictions and do not rely on applying an assumed fixed flux ratio.
   
Our results highlight the importance of considering effects such as the finite beam-size of single-dish telescopes and field-to-field variance when comparing sub-mm observations with theoretical models.  In our model SMGs are predominantly a disc instability triggered starburst population, the sub-mm emission of which is often blended along the line of sight of observational single-dish surveys.  

In future work we will conduct a more thorough investigation of the properties and evolution of SMGs within the model presented in L14, including an analysis of their clustering with and without the effects of the single-dish beam.  We hope that this, when compared with future observations aided by sub-mm interferometry of increasing sample sizes, will lead to a greater understanding of this extreme and important galaxy population. 

\section*{Acknowledgments}
The authors would like to thank James Simpson and Chian-Chou Chen for helpful discussions.  We also thank the anonymous referee for a detailed and constructive report which allowed us to improve the quality of the manuscript.  This work was supported by the Science and Technology Facilities Council [ST/K501979/1, ST/L00075X/1].  This work used the DiRAC Data Centric system at Durham University, operated by the Institute for Computational Cosmology on behalf of the STFC DiRAC HPC Facility (www.dirac.ac.uk). This equipment was funded by BIS National E-infrastructure capital grant ST/K00042X/1, STFC capital grant ST/H008519/1, and STFC DiRAC Operations grant ST/K003267/1 and Durham University. DiRAC is part of the National E-Infrastructure. 

\bibliographystyle{mn2e}
\bibliography{ref.bib}

\appendix

\label{lastpage}

\end{document}